\documentclass[usenatbib]{mn2e} 
\usepackage{longtable} 
\usepackage{graphicx}

\title[Study of open cluster Berkeley 24 and Czernik 27]
      {Mass function and dynamical study of the open clusters Berkeley 24 and Czernik 27}

\author[Bisht et al.]
   {D. Bisht$^{1,3}$,\thanks{E-mail: dbisht@prl.res.in; rkant@aries.res.in; shashi@prl.res.in; alokdurgapal@gmail.com; geetarangwal91@gmail.com}
R.\ K.\ S.\ Yadav$^{2}$,   
Shashikiran Ganesh$^{3}$,  
A.\ K. Durgapal$^{4}$,    
G.\ Rangwal$^{4}$, and
\and
J.\ P.\ U.\ Fynbo$^{5}$   
\\\\
    $^{1}$ Key Laboratory for Researches in Galaxies and Cosmology, University of Science and Technology of 
           China,\\ 
           Chinese Academy of Sciences, Hefei, Anhui, 230026, China \\ 
    $^{2}$Aryabhatta Research Institute of Observational Sciences, 
           Manora Peak, Nainital 263129, India\\
    $^{3}$Physical Research Laboratory, Ahmedabad, 380009, India \\ 
    $^{4}$Department of physics, DSB campus, KU Nainital, India\\
    $^{5}$The Cosmic Dawn Center, Niels Bohr Institute, Copenhagen University, Juliane Maries Vej 30, DK-2100 Copenhagen {\O}
}

\begin{document}

\date{\today}


\maketitle

\label{firstpage}

\begin{abstract}

We present a $UBVI$ photometric study of the open clusters Berkeley 24 (Be 24) and Czernik 27 (Cz 27). The radii of the clusters 
are determined as 2\farcm7 and 2\farcm3 for Be 24 and Cz 27, respectively. We use the Gaia 
Data Release 2 (GDR2) catalogue to estimate the mean proper motions for the clusters. We found the mean proper motion of 
Be 24 as $0.35\pm0.06$ mas yr$^{-1}$ and $1.20\pm0.08$ mas yr$^{-1}$ in right ascension and declination for Be 24 and 
$-0.52\pm0.05$ mas yr$^{-1}$ and $-1.30\pm0.05$ mas yr$^{-1}$ for Cz 27. We used probable cluster members selected from proper
motion data for the estimation of fundamental parameters. We infer reddenings $E(B-V)$ = $0.45\pm0.05$ mag
and $0.15\pm0.05$ mag for the two clusters. Analysis of extinction curves towards the two clusters show that both have normal interstellar
extinction laws in the optical as well as in the near-IR band. From the ultraviolet excess measurement, we derive metallicities of 
[Fe/H]= $-0.025\pm0.01$ dex and $-0.042\pm0.01$ dex for the clusters Be 24 and Cz 27, respectively. The distances, as determined from 
main sequence fitting, are $4.4\pm0.5$ kpc and $5.6\pm0.2$ kpc. The comparison of observed CMDs with $Z=0.01$ isochrones, leads to an
age of $2.0\pm0.2$ Gyr and $0.6\pm0.1$ Gyr for Be 24 and Cz 27, respectively. 

In addition to this, we have also studied the mass function and dynamical state of these two clusters for the first time using probable
cluster members. The mass function is derived after including the corrections for data incompleteness and field star contamination. 
Our analysis shows that both clusters are now dynamically relaxed. 

\end{abstract}



\section{Introduction} \label{sec:intro}

  \begin{figure*}
    \centering
   \hbox{ 
   \includegraphics[width=9cm,height=9cm]{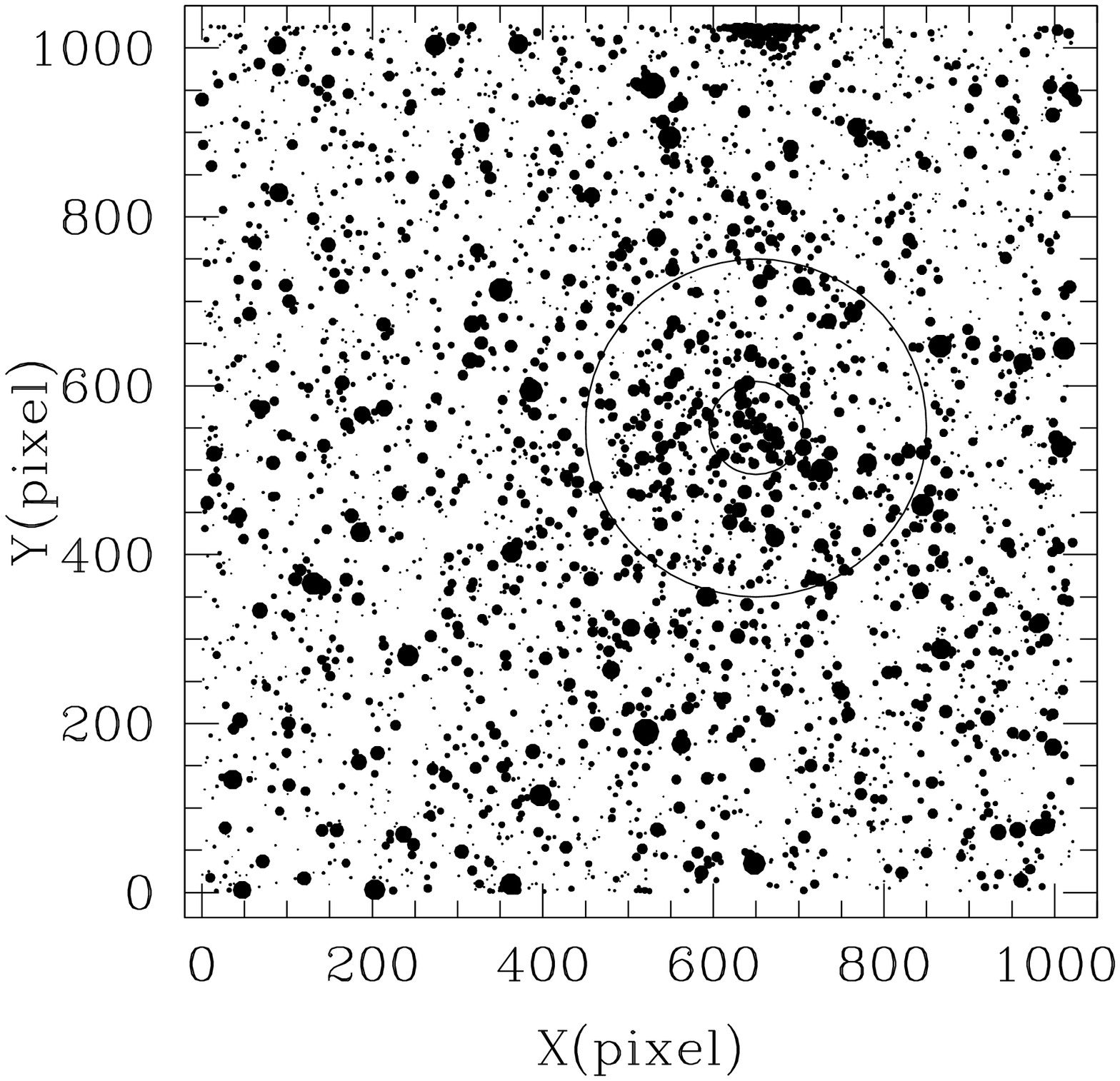}
    
 \includegraphics[width=9cm,height=9cm]{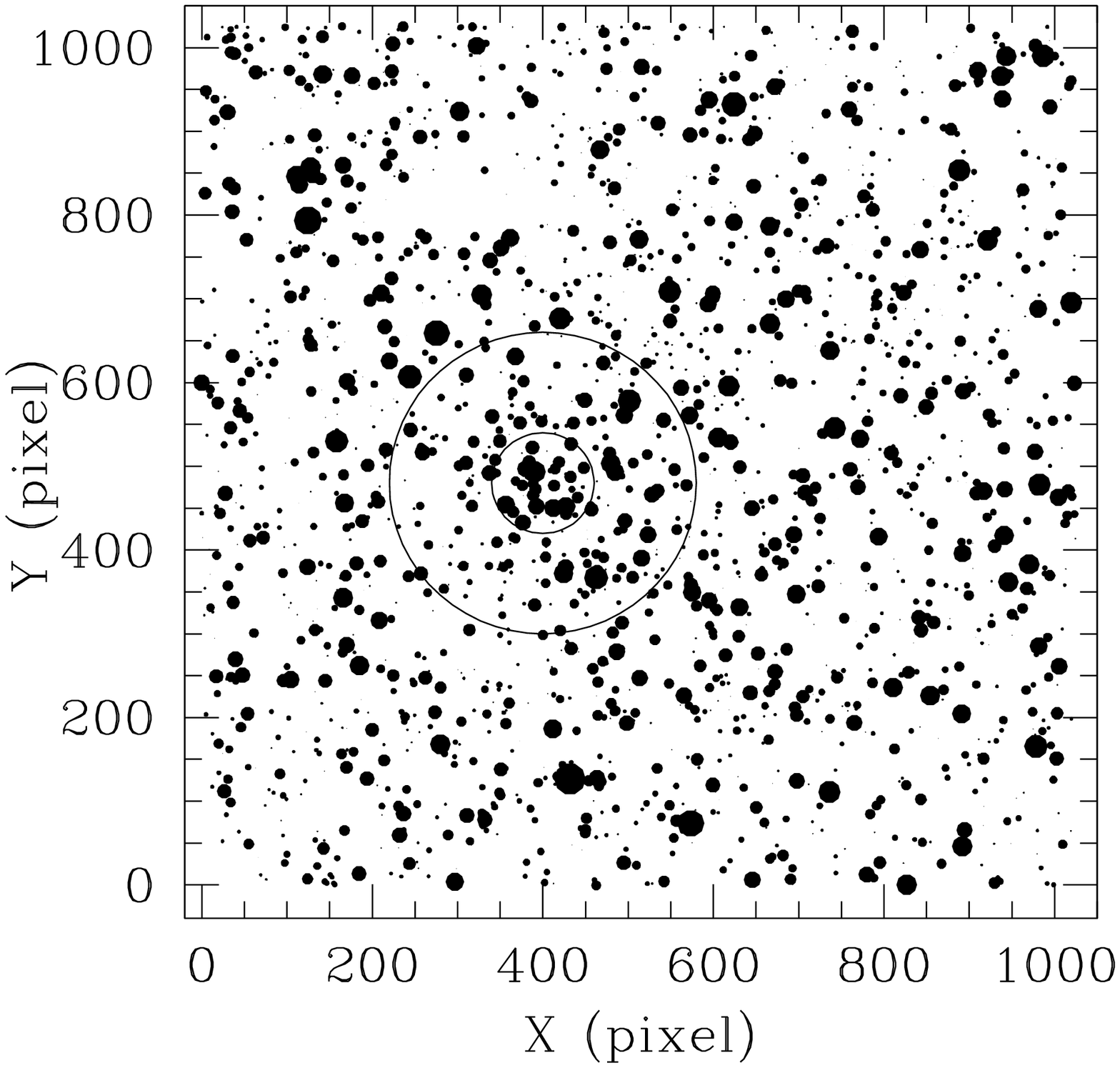}
}
\caption{Finding chart of the stars in the field of Be 24 and Cz 27. 
Filled circles of different sizes represent brightness of the stars. Smallest size 
denotes stars of $V$$\sim$20 mag. Open outer circle represent the cluster size and 
inner circle represent core region.}
  \label{chart}
  \end{figure*}

Galactic clusters contain from a few tens to several thousands stars, which are loosely concentrated and gravitationally bound 
to each other. Open clusters are important tools for the study of Galactic structure (Janes \& Alder 1982), chemical evolution 
(Magrini et al. 2009) and star formation processes in the Milky Way Galaxy. Galactic clusters offer the advantage of studies of
field stars that their ages and distances can be determined from main sequence fitting (Carraro et al. 2013). In particular, the 
intermediate/old age open clusters are very useful in testing stellar isochrones and dynamical evolution of the stars. Furthermore,
they are found in all parts of Milky Way galaxy and exhibit very wide range in their ages. A new era in dynamical astronomy
has begun with the second data release of the Gaia (GDR2) mission in April 2018 (Brown et al. 2018).

The distribution of stellar masses that results from a star formation event is called the Initial Mass Function (IMF). The IMF is 
the key parameter for constraining star formation theories and also to understand the subsequent stellar evolution process. In the
recent years, the mass function has been determined for a number of open clusters by numerous authors (e.g., Hur et al. 2012; 
Melnikov et al. 2012; Khalaj et al. 2013). It is still not well understood if the shape of the IMF is universal in time and space or
if it depends upon astrophysical parameters like metallicity, cluster extent, etc. (e.g., Scalo 1986, 1998). 

\begin{table*}
\centering
\caption{Fundamental parameters of the clusters Be 24 and Cz 27 are taken from Dias et al. (2002) and WEBDA.}
\vspace{0.5cm} 
\begin{tabular}{ccccccccc}
\hline\hline
Name       & $\alpha_{2000}$  & $\delta_{2000}$ & $l$& $b$& Dia & $E(B-V)$     &
$D$   & log(age)   \\
           & h:m:s            & d:m:s         & (deg) &(deg)  & ($\prime$) &   (mag)        &
 (pc)       &            \\
\hline
Be 24 & 06:37:47       & $-$00:52:19       & 210.6 & $-$2.6 &   7     &0.4& 4700 &9.34\\
Cz 27 & 07:03:22       & 6:23:47       & 208.5 & 5.5 &   5     &0.15& 5800&8.80\\
\hline
\end{tabular}
\label{astro_para}
\end{table*}

Open clusters are very important objects for the investigation of the dynamical evolution of stellar system, because the stars 
in clusters are born from the same molecular cloud and have evolved in the same gravitational potential. Energy equipartition 
due to interactions between cluster stars leads to mass segregation. The equipartition of 
energy takes place after many encounters between the members of a cluster. This causes massive stars to lose kinetic
energy and fall toward the cluster center, while lower mass stars increase their velocity and move outward. The encounters 
continue until the system becomes relaxed. The relaxation time scale depends on the number of stars in a cluster 
(Bonnell \& Davies, 1998). Mass segregation results in a steeper mass function slope in the outer region of the cluster 
(Ann \& Lee 2002). The process also causes some fraction of the low mass stars to evaporate from the cluster to create a halo 
of predominantly low mass stars surrounding the cluster (Eggen 1993).   

In the present study, we consider two open clusters, Berkeley 24 and Czernik 27 (Be 24 and Cz 27 in the following),
which are situated between Perseus and the outer arm in the third Galactic quadrant of the Milky Way. The basic parameters
available in the literature (Dias et al. 2002) are provided in Table~\ref{astro_para}. Here we list what has been 
published on these clusters in the literature:

Be 24: Ortolani et al. (2005) studied this cluster using $BV$ photometry and 
estimated the reddening and distance to be $E(B-V)=0.40$ mag and 4.7 kpc, respectively. 

Cz 27: This is a relatively faint cluster recognized in Monoceros by Czernik 
(1966). Kim et al. (2005) found that Cz 27 is a moderately reddened $(E(B-V)=0.15)$ cluster with an age similar to the 
Hyades located 5.8 kpc from the Sun. Cz 27 was later studied by Piatti et al. (2010) using $UBVI_{KC}$ 
photometry and they found that this is an intermediate age star cluster. They derived the reddening to be $E(B-V)=0.08$ mag,
and inferred a significantly smaller distance of 2.1 kpc and an age of about 0.7 Gyr.

From the previous studies, we can see that there is some disagreement in the basic parameters of the clusters
between different studies. Hence, a detailed study of these objects is justified to try to clarify these disagreements. 
In addition to this, these clusters make a bridge between young open clusters and globular clusters in understanding the dynamical 
evolution of clusters. Therefore, in the present article, we provide new $UBVI$ CCD photometry of Be 24 and Cz 27, and study
their basic parameters along with their mass functions and dynamical evolution. 

The plan of this paper is as follows. We describe our observations and data reduction techniques in Section 2.
Section 3 and 4 deal with the derivation of the basic parameters of the clusters. 
Section 5 is devoted to the luminosity and mass function of the clusters. The dynamical state 
of the clusters is described in Section 6. Finally, we summarize our results in the last section.

\begin{table}
\begin{center}
\caption{Log of observations, with dates and exposure times for each passband.}
\vspace{0.5cm}
\begin{tabular}{ccc}
\hline\hline
Band  &Exposure Time &Date\\
&(in seconds)   & \\
\hline\hline
&Be 24&\\
$U$&1500$\times$2, 300$\times$1&5$^{th}$ Dec 2010 \\
$B$&1200$\times$2, 240$\times$2&,,\\
$V$&900$\times$3, 120$\times$1&,,\\
$I$&300$\times$2, 120$\times$2&,,\\
\hline
&Cz 27&\\
$U$&1500$\times$2, 300$\times$1&7$^{th}$ Nov 2010 \\
$B$&1200$\times$2, 240$\times$2&,,\\
$V$&900$\times$2, 180$\times$2&,,\\
$I$&300$\times$2, 120$\times$1&,,\\
\hline
\end{tabular}
\label{log}
\end{center}
\end{table}

\section{Observations and data reduction}

We have used CCD imaging to obtain new $UBVI$ photometry of stars in the region of our two clusters of interest, 
Be 24 and Cz 27.
These data were obtained using the 104-cm Sampurnanand reflector telescope (f/13) located at Aryabhatta Research 
Institute of Observational Sciences, Manora Peak, Nainital, India. Images were acquired using a 2K$\times$2K CCD,
which has 24 $\mu$m square pixels, 
resulting in a scale of 0$^{\prime\prime}$.36 pixel$^{-1}$ and a square field of view of 12.$^{\prime}$6
size. The CCD gain was 10 e$^{-}$/ADU while the read out noise was 5.3 e$^{-}$. In order to improve the S/N ratio,
the observations were taken in the 2$\times$2 pixel$^2$ binned mode. The observations were organised in several
short exposures in each of the filters as specified in Table~\ref{log}. In Table~\ref{log} we also list the 
observing dates. The identification maps based on our $V$-band observations are shown in Fig~\ref{chart}.

To clean the science images, a number of bias and twilight flat-field frames were taken in $V$, $B$, $I$ and
$U$, during the two observing nights. The IRAF\footnote{ IRAF is distributed by the National Optical Astronomical Observatory which are 
operated by the Association of Universities for Research in Astronomy, under contract with the National Science 
Foundation} data reduction package was used for initial processing of data frames which includes bias 
subtraction, flat fielding and cosmic ray removal. Stellar magnitudes were obtained by using the DAOPHOT 
software. The instrumental magnitudes were derived through point spread function (PSF) fitting using DAOPHOT/ALLSTAR (Stetson 
1987, 1992) package. To determine the PSF, we used several well isolated stars distributed over the entire frame. 
The Gaussian function 
was used as an analytical model PSF. The shape of the PSF was made to vary quadratically with position on the frame. 
Appropriate aperture corrections were calculated using isolated and unsaturated bright stars in the frame. 

ALLSTAR computes $x$ and $y$ centers, sky values and magnitudes for the stars by fitting the PSF to groups of stars 
in the image. Initial estimates of the centers, sky values and magnitudes are read from the aperture photometry file.
ALLSTAR groups the stars dynamically, performing a regrouping operation after every iteration. The new computed 
centers, sky values and magnitudes are written in a file along with the number of iterations it took to fit the star,
the goodness of fit (chi) and sharpness. An image with all the fitted stars subtracted out is written in another image.
In effect ALLSTAR performs the combined operations of GROUP, NSTAR and SUBSTAR from Daophot.

We have cross-identified the stars of different frames and filters using the DAOMATCH/DAOMASTER programme available in 
DAOPHOT II. To determine the transformation coefficients from instrumental to standard magnitudes the CCDLIB and CCDSTD 
routines were used. Finally, standard magnitudes and colours of all the stars were obtained using the routine FINAL.  

\begin{table}
\begin{center}
\caption{Derived Standardization coefficients and their errors.}
\vspace{0.5cm}
\begin{tabular}{ccc}
\hline\hline
Filter  &   Colour Coeff. $(C)$ & Zeropoint $(Z)$\\
\hline\hline
&Be 24&\\ 

$U$&$-0.040\pm$0.09&$7.35\pm$0.05\\
$B$&$+0.001\pm$0.01&$5.29\pm$0.01\\
$V$&$-0.104\pm$0.01&$4.99\pm$0.01\\
$I$&$-0.125\pm$0.01&$5.43\pm$0.01\\

&Cz 27&\\

$U$&$-0.03\pm$0.02&$7.85\pm$0.01\\
$B$&$-0.03\pm$0.01&$5.64\pm$0.01\\
$V$&$-0.06\pm$0.01&$5.21\pm$0.01\\
$I$&$-0.06\pm$0.01&$5.48\pm$0.01\\
\hline
\end{tabular}
\label{c_coff}
\end{center}
\end{table}

\begin{table}
\centering
\caption{The rms global photometric errors as a function of $V$ magnitude.}
\vspace{0.5cm}
\begin{tabular}{ccccc}
\hline
$V$&$\sigma_{V}$&$\sigma_{B}$&$\sigma_{I}$&$\sigma_{U}$ \\
\hline
$10-11$&$0.05$&$0.05$&$0.05$&$0.05$ \\
$11-12$&$0.05$&$0.05$&$0.03$&$0.05$ \\
$12-13$&$0.05$&$0.05$&$0.05$&$0.06$ \\
$13-14$&$0.05$&$0.05$&$0.05$&$0.06$ \\
$14-15$&$0.05$&$0.05$&$0.05$&$0.06$ \\
$15-16$&$0.05$&$0.05$&$0.05$&$0.08$ \\
$16-17$&$0.05$&$0.05$&$0.05$&$0.11$ \\
$17-18$&$0.06$&$0.06$&$0.06$&$0.13$ \\
$18-19$&$0.06$&$0.06$&$0.06$&$0.20$ \\
$19-20$&$0.07$&$0.08$&$0.07$&$0.31$ \\
\hline
\end{tabular}
\label{g_error}
\end{table}

\subsection {Photometric calibration}

We observed the standard field SA 98 (Landolt 1992) during both observing nights
for the purpose of photometric calibration of the CCD system. The standard stars used in the calibrations 
have brightness and colour range 12.77 $\le V \le$ 16.11 and $-0.329 < (B-V) < 1.448$, respectively,
thus covering the range relevant for the bulk of the cluster stars. For the atmospheric extinction coefficients 
we assumed the typical values 
for the ARIES site (Kumar et al. 2000). For translating the instrumental magnitude to the standard magnitude, 
the calibration equations derived using least square linear regression are as follows:

\begin{center}
   $u=U+Z_{U}+C_{U}(U-B)+k_{U}X$\\

   $b=B+Z_{B}+C_{B}(B-V)+k_{B}X$\\

   $v=V+Z_{V}+C_{V}(B-V)+k_{V}X$\\

   $i=I+Z_{I}+C_{I}(V-I)+k_{I}X$
\end{center}

where $u, b, v$ and $i$ are the aperture instrumental magnitudes, $U, B, V$ and $I$ are
the standard magnitudes, and $X$ is the airmass. The colour coefficients (C) and zeropoints (Z) for
the different filters are listed in Table~\ref{c_coff}. The errors in zero points and colour coefficients
are $\sim$0.01 mag except in the $U$ filter where it is 0.09 mag. The internal errors derived from 
DAOPHOT are plotted against $V$ magnitude in Fig.~\ref{error}. This figure shows that the average photometric 
error is $\le$ 0.01 mag for $B$, $V$ and $I$ filters at $V\sim19^{th}$ mag, while it is $\le$ 0.03 mag for $U$ 
filter at $V\sim18^{th}$ mag. Global photometric (DAOPHOT+Calibrations) errors are also calculated and
listed in Table~\ref{g_error}. For the $V$ filter, the errors are 0.05 at $V\sim$17 mag and 0.07 at $V\sim$20 mag. 
The final photometric data are available in electronic form at the WEBDA site \footnote{\it http://obswww.unige.ch/webda/}
and also upon request directly from the authors.

In order to transform CCD pixel coordinates to celestial coordinates, we have used the online digitized European 
Southern Observatory catalogue included in the $SKYCAT$ software as an absolute astrometric reference frame. The 
$CCMAP$ and $CCTRAN$ routines in $IRAF$ were used to find a transformation equation which gives the celestial coordinates 
as a function of the pixel coordinates. The resulting celestial coordinates have a standard deviation of 0.1 arcsec in 
both right ascension and declination.

\subsection{\bf The 2MASS data}

\begin{figure}
    \centering
\hbox{
   \includegraphics[width=4.5cm, height=4cm]{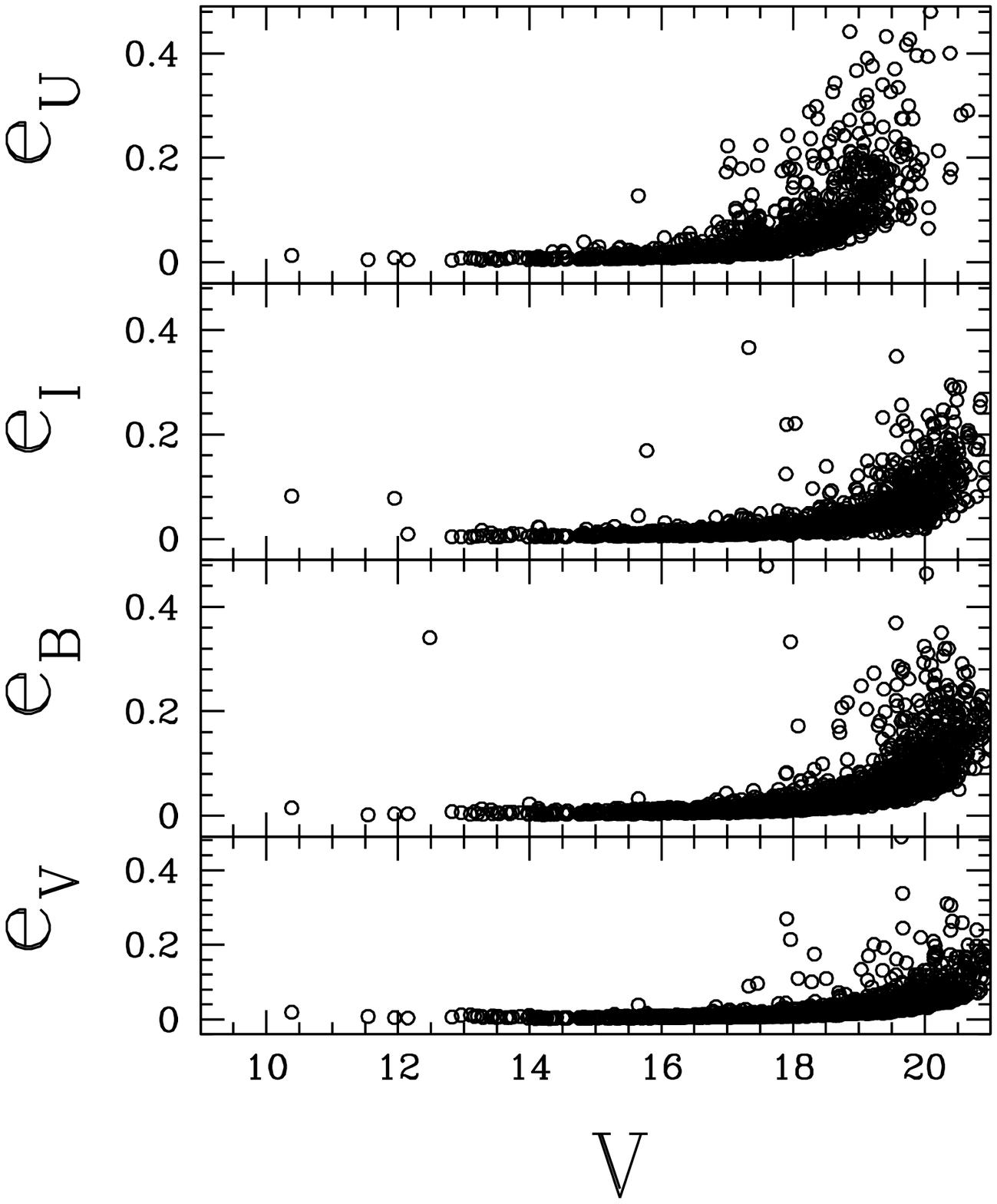}
   \includegraphics[width=4.5cm, height=4cm]{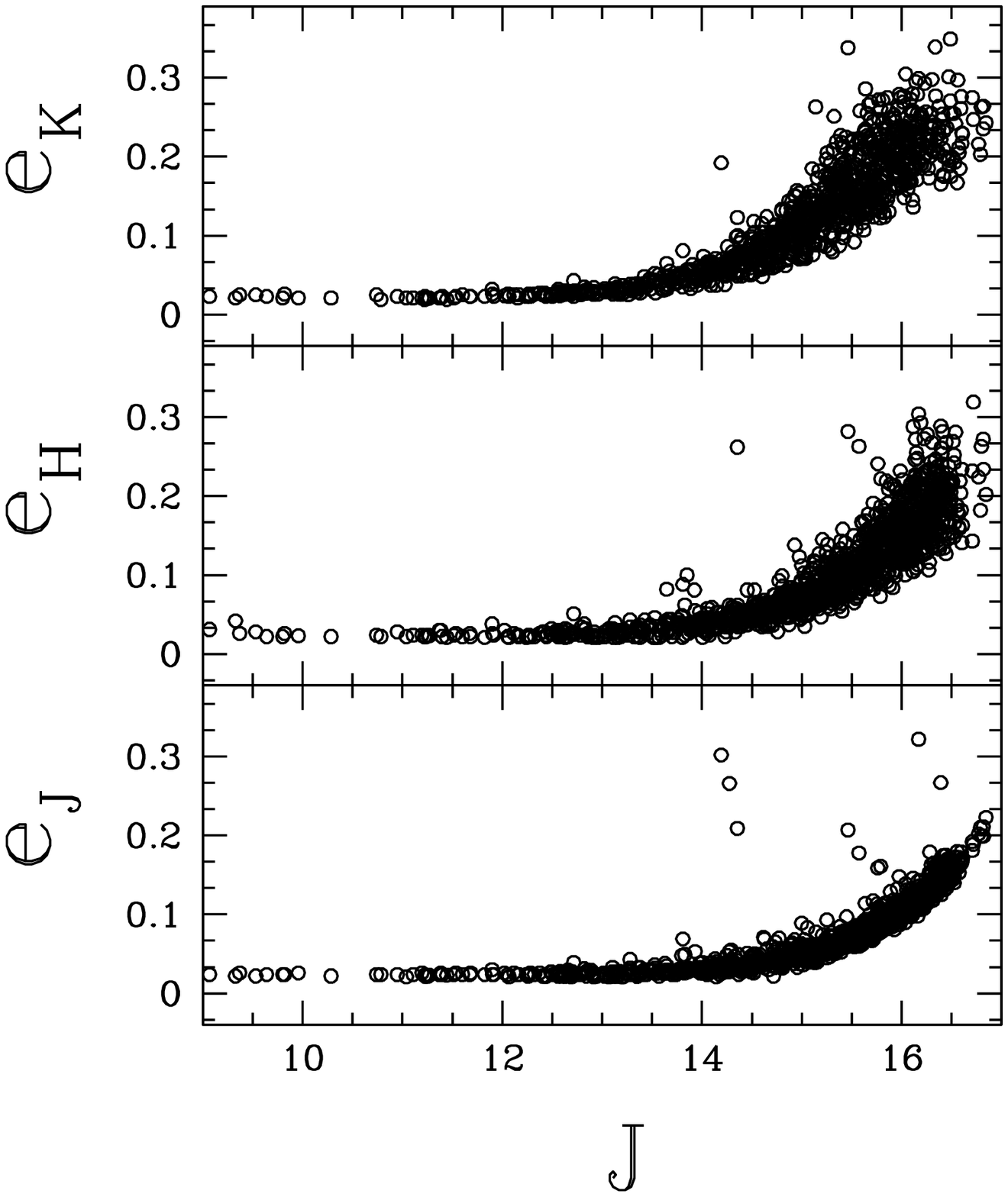}
}
\caption{Photometric errors in different bands against $V$ magnitude (Left) and $J$ magnitude (right).}
  \label{error}
  \end{figure}

The near-Infrared $JHK_{s}$ photometric data for clusters Be 24 and Cz 27 were taken from the Two Micron
All-sky Survey (2MASS). $2MASS$ uniformly scanned the entire sky in three near-IR bands $J (1.25\micron)$,
$H (1.65\micron)$ and $K_{s} (2.17\micron)$. The 2MASS (Skrutskie et al. 2006) used two highly automated 
1.3m aperture, open tube, equatorial fork-mount telescopes (one at Mt. Hopkins, Arizona (AZ), USA and 
other at CTIO, Chile) with a 3-channel camera $(256\times256)$ array of HgCdTe detectors in each channel).
The $2MASS$ data base provides photometry in the near infrared $J$, $H$ and $K_{s}$ bands to a limiting 
magnitude of 15.8, 15.1 and 14.3 respectively, with a signal to noise ratio (S/N) greater than 10. We retain
only those sources for which the error in each band is less than 0.15 mag to ensure a sufficient photometric 
accuracy. The errors given in the $2MASS$ catalogue for the $J$, $H$ and $K_{s}$ bands are plotted against $J$ magnitudes
in Fig.~\ref{error}. This figure shows that the mean errors in the $J$, $H$ and $K_{s}$ bands are all $\le0.05$ mag
at $J\sim$13 mag. The errors become $\sim$0.1 mag at $J\sim$15 mag. 

  \begin{figure}
    \centering
\hbox{
   \includegraphics[width=4.5cm, height=6cm]{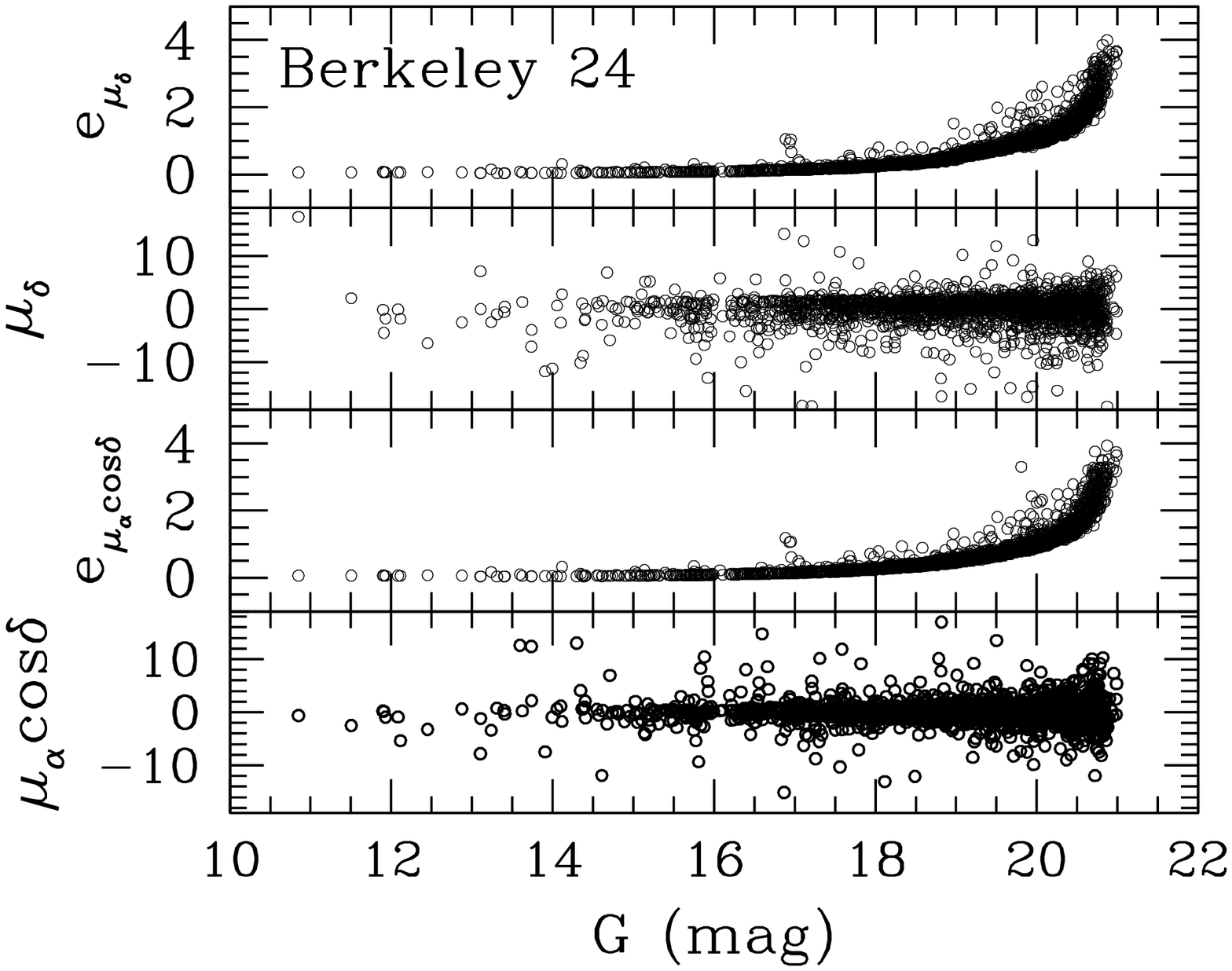}
   \includegraphics[width=4.5cm, height=6cm]{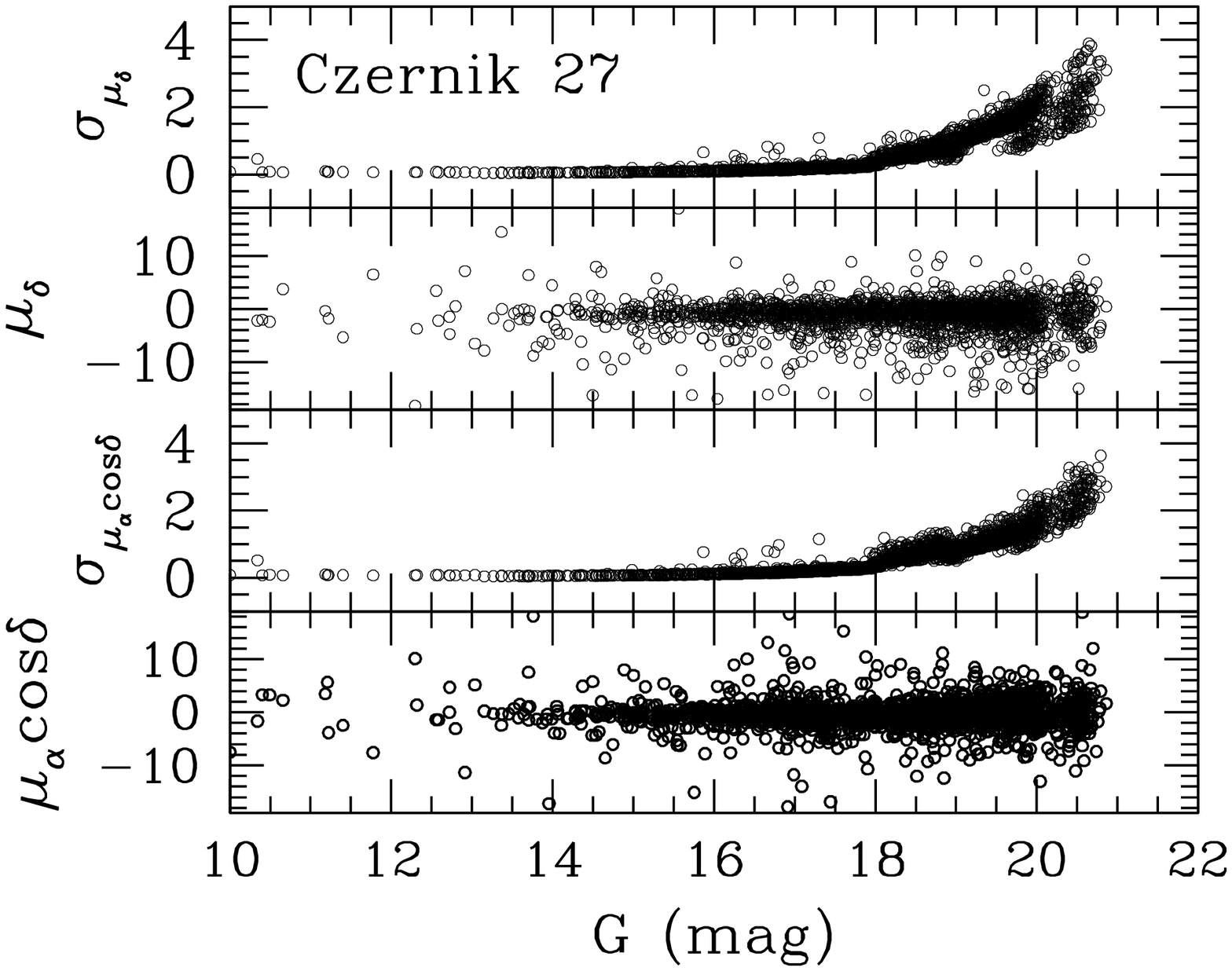}
}
\caption{Plot of proper motions and their errors versus $G$ magnitude.}
  \label{error_pm}
  \end{figure}

  \begin{figure}
    \centering
    \includegraphics[width=8cm, height=8cm]{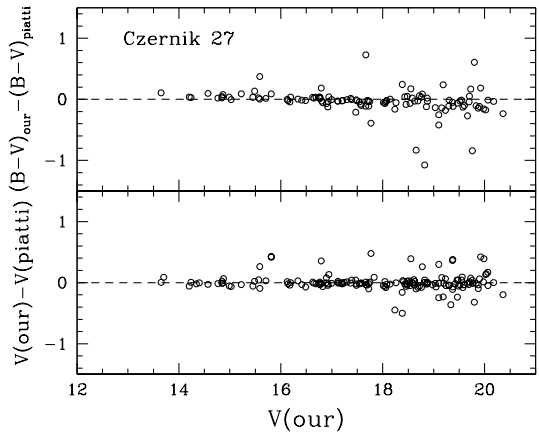}
\caption{Differences between measurements presented in Piatti et al. (2010) and in the present study for
$V$ magnitude and $(B-V)$ colours. Zero difference is indicated by the dashed line.}
  \label{match_p}
  \end{figure}

\subsection{\bf Gaia DR2}

We used GDR2 (Brown et al. 2018) for proper motion study of clusters Be 24 and Cz 27. This data contains
five parametric astrometric solution-positions on the sky $(\alpha, \delta)$, parallaxes and ($\mu_{\alpha} cos\delta ,
\mu{\delta}$) with a limiting magnitude of $G=21$ mag. Parallax uncertainties are in the range of up to 0.04 milliarcsecond (mas)
for sources at $G\le15$ mag, around 0.1 mas for sources with $G\sim17$ mag. The uncertainties in the respective proper motion
components are up to 0.06 mas $yr^{-1}$ (for $G\le15$ mag), 0.2 mas $yr^{-1}$ (for $G\sim17$ mag) and 1.2 mas $yr^{-1}$ 
(for $G\sim20$ mag). The proper motion and their corresponding errors are plotted against $G$ magnitude as shown in 
Fig \ref{error_pm} for clusters Be 24 and Cz 27. In this figure errors in proper motion components are $\sim 1.2$ at $G\sim20$ mag.



\subsection{Comparison with previous photometry}

\begin{table}
\centering
\caption{Differences in $V$ and $(B-V)$ between Piatti et al. (2010) and our study. The standard
deviation in the difference for each magnitude bin is also given in the parentheses.}
\vspace{0.5cm}
\begin{tabular}{crr}
\hline
$V$&$\Delta{V}$&$\Delta(B-V)$ \\
\hline
$13-14$&$-0.02~ (0.03)$&$0.02~ (0.04)$ \\
$14-15$&$-0.07~ (0.05)$&$0.03~ (0.03)$ \\
$15-16$&$0.06~ (0.20)$&$0.01~ (0.12)$ \\
$16-17$&$-0.06~ (0.08)$&$-0.09~ (0.18)$ \\
$17-18$&$-0.05~ (0.10)$&$-0.07~ (0.26)$ \\
$18-19$&$-0.09~ (0.14)$&$-0.04~ (0.18)$ \\
$19-20$&$-0.04~ (0.19)$&$-0.11~ (0.25)$ \\
\hline
\end{tabular}
\label{match_error}
\end{table}

The $CCD$ $UBVI_{KC}$ photometry down to $V\sim$21.0 for the open cluster Cz 27 has been discussed by 
Piatti et al. (2010). We have cross-identified stars in the two catalouges on the assumption that stars are 
correctly matched if the difference in position is less than 1 arcsec. On this basis, we have found 154 common
stars. A comparison of $V$ magnitudes and $(B-V)$ colours
between the two catalouges is shown in Fig.~\ref{match_p}. The mean difference and standard deviation per 
magnitude bin are given in Table~\ref{match_error}. This indicates that our $V$ and $(B-V)$ measurements are 
in fair agreement with those given in Piatti et al. (2010). For Be 24, there is no photometric 
catalouges available in the literature. Our deep photometry  for this cluster is the first publicly 
available catalogue.


\section{Mean Proper motion of the clusters}

To derive the mean proper motion of the clusters, $GDR2$ data are used for both the clusters. In this catalog we can 
find the mean positions and proper motions for all objects down to $G\sim$ 20 mag.

\begin{figure*}
\begin{center}
\centering
\hbox{
\hspace{1cm}\includegraphics[width=8.0cm, height=8.0cm]{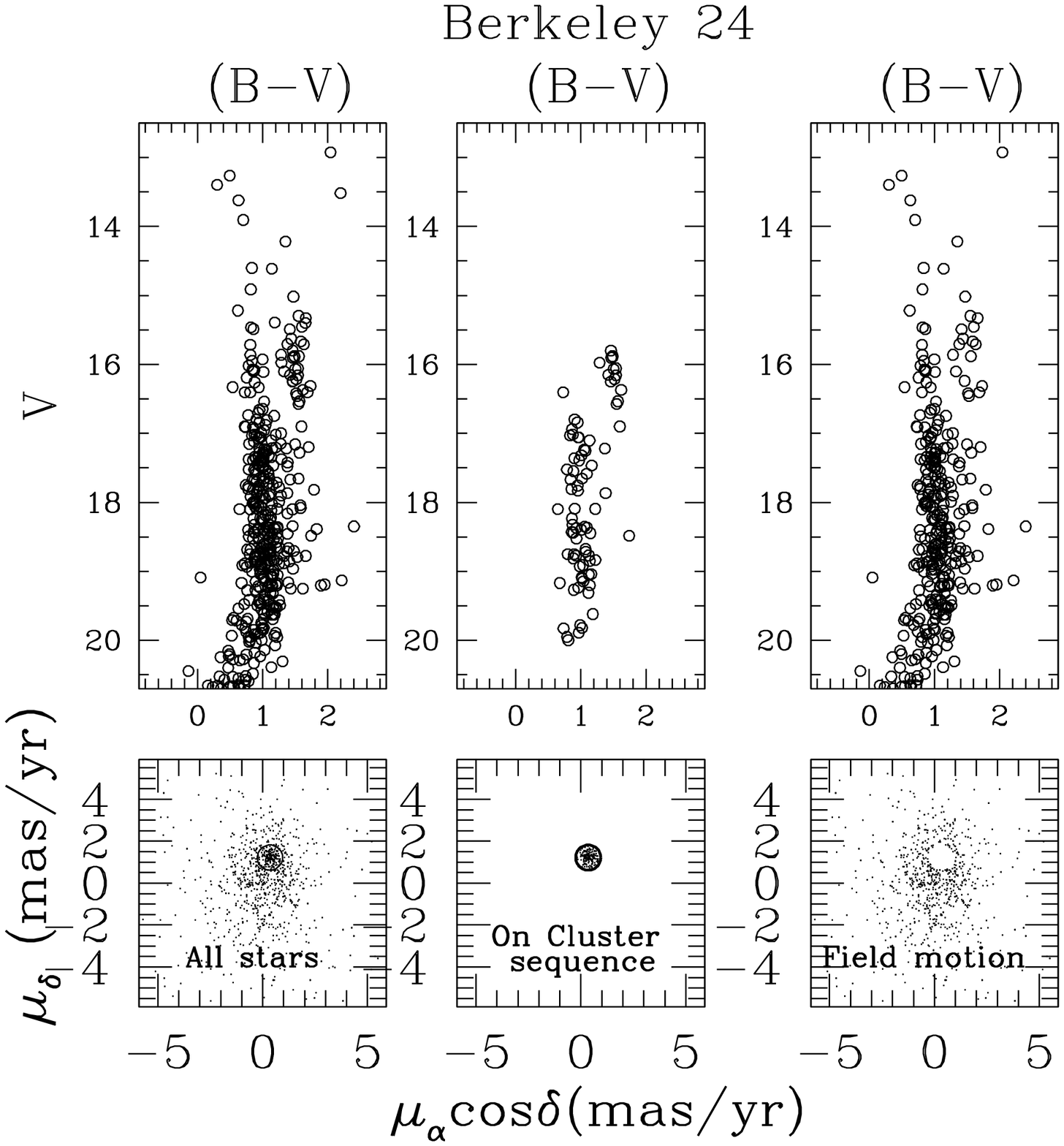}
\includegraphics[width=8.0cm, height=8.0cm]{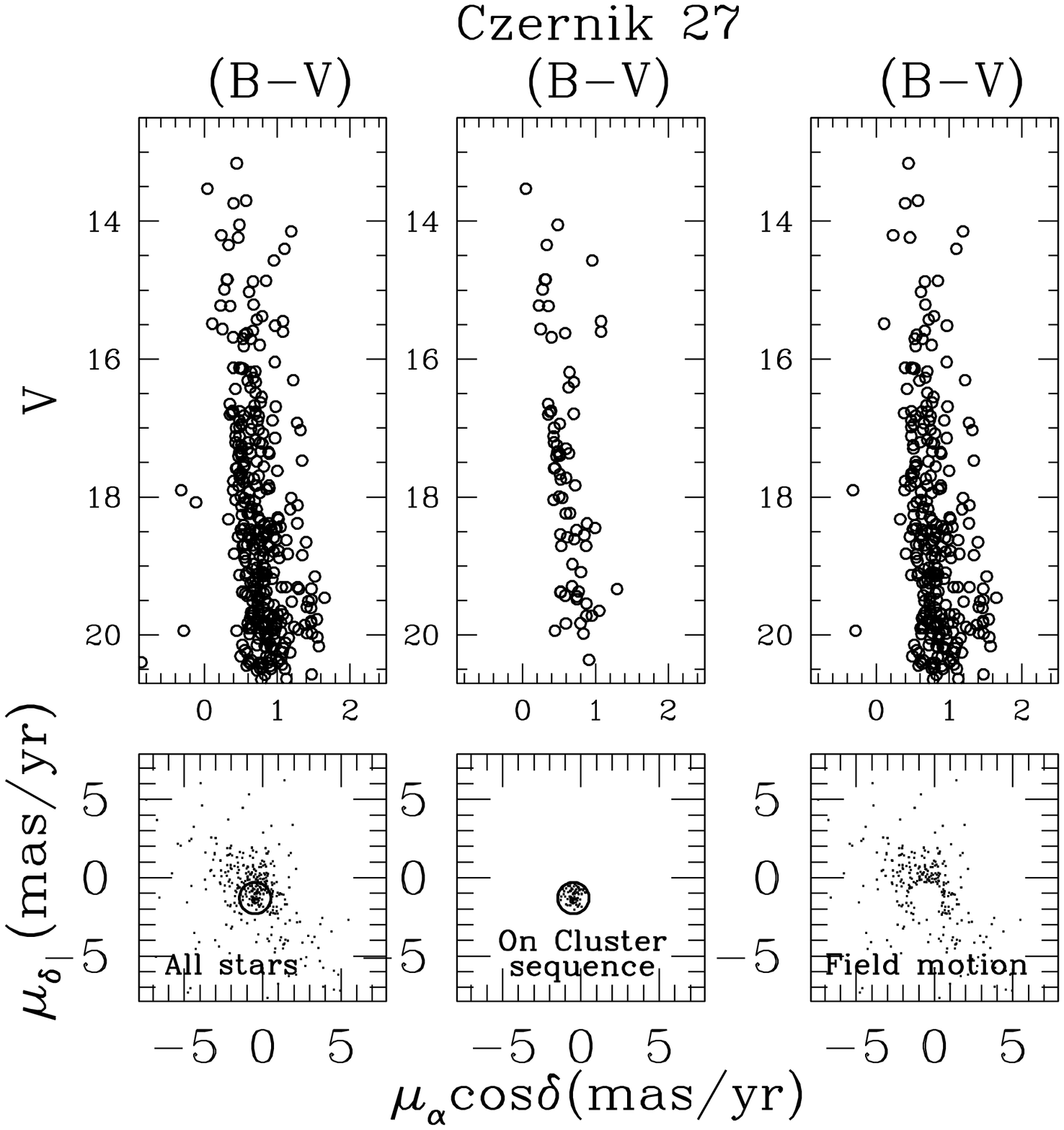}
}
\caption{(Bottom panels) Proper-motion vector point diagram (VPD) for Be 24 and Cz 27. (Top panels)
$V$ versus $(B-V)$ colour magnitude diagrams. (Left) The entire sample. (Center) Stars in VPDs within circle of $0.6~ mas~ yr^{-1}$
and $1~ mas~ yr^{-1}$ for Be 24 and Cz 27 respectively of the cluster mean. (Right) Probable background/foreground field stars in the
direction of these clusters. All plots show only stars with PM $\sigma$ smaller than $1~ mas~ yr^{-1}$ in each coordinate.} 
\label{pm_cmd}
\end{center}
\end{figure*}

The mean proper motion is defined as the average angular speed of cluster per year by which it has changed its position 
over the sky. PMs $\mu_{\alpha} cos{\delta}$ and $\mu{\delta}$ are plotted as VPDs in the bottom panels of 
Fig \ref{pm_cmd}. The top panels shows corresponding $V$ versus $(B-V)$ CMDs. The left panel show all stars, while the middle 
and right panel show the probable cluster members and field stars. A circle of 0.6 and 1 mas $yr^{-1}$ around the cluster center 
for Be 24 and Cz 27 respectively in the VPD defines our membership criteria. The chosen radius is a compromise between loosing 
cluster members with poor PMs and including field region stars. The CMD of the probable cluster members are shown in the 
upper-middle panels in Fig \ref{pm_cmd}. The main sequence of the clusters are clearly separated out. These stars have a 
PM error of $\le 1 ~ mas yr^{-1}$. 

\begin{figure}
\begin{center}
\centering
\hbox{
\includegraphics[width=4.0cm,height=4.0cm]{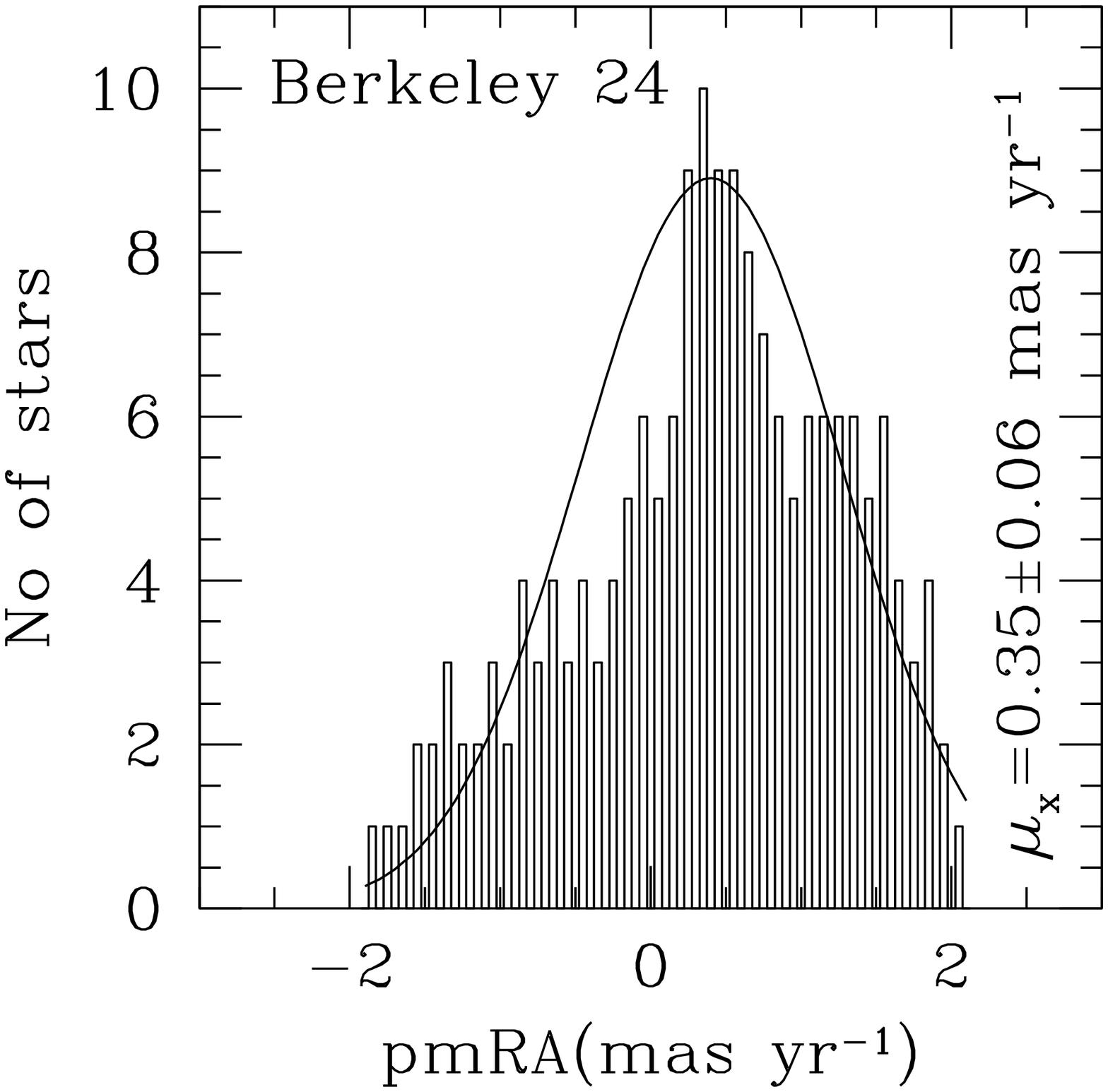}
\includegraphics[width=4.0cm,height=4.0cm]{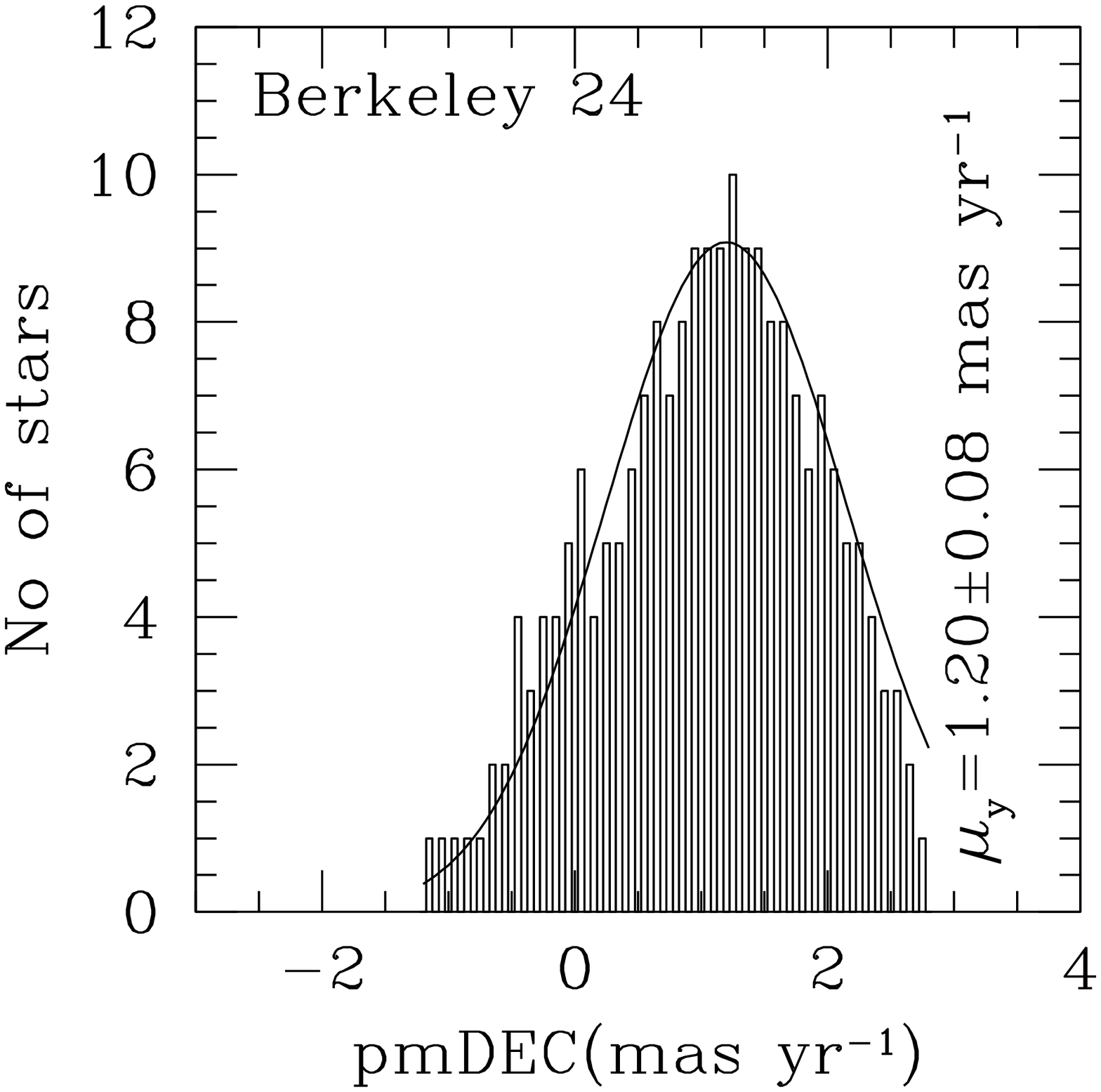}
}
\hbox{
\includegraphics[width=4.0cm,height=4.0cm]{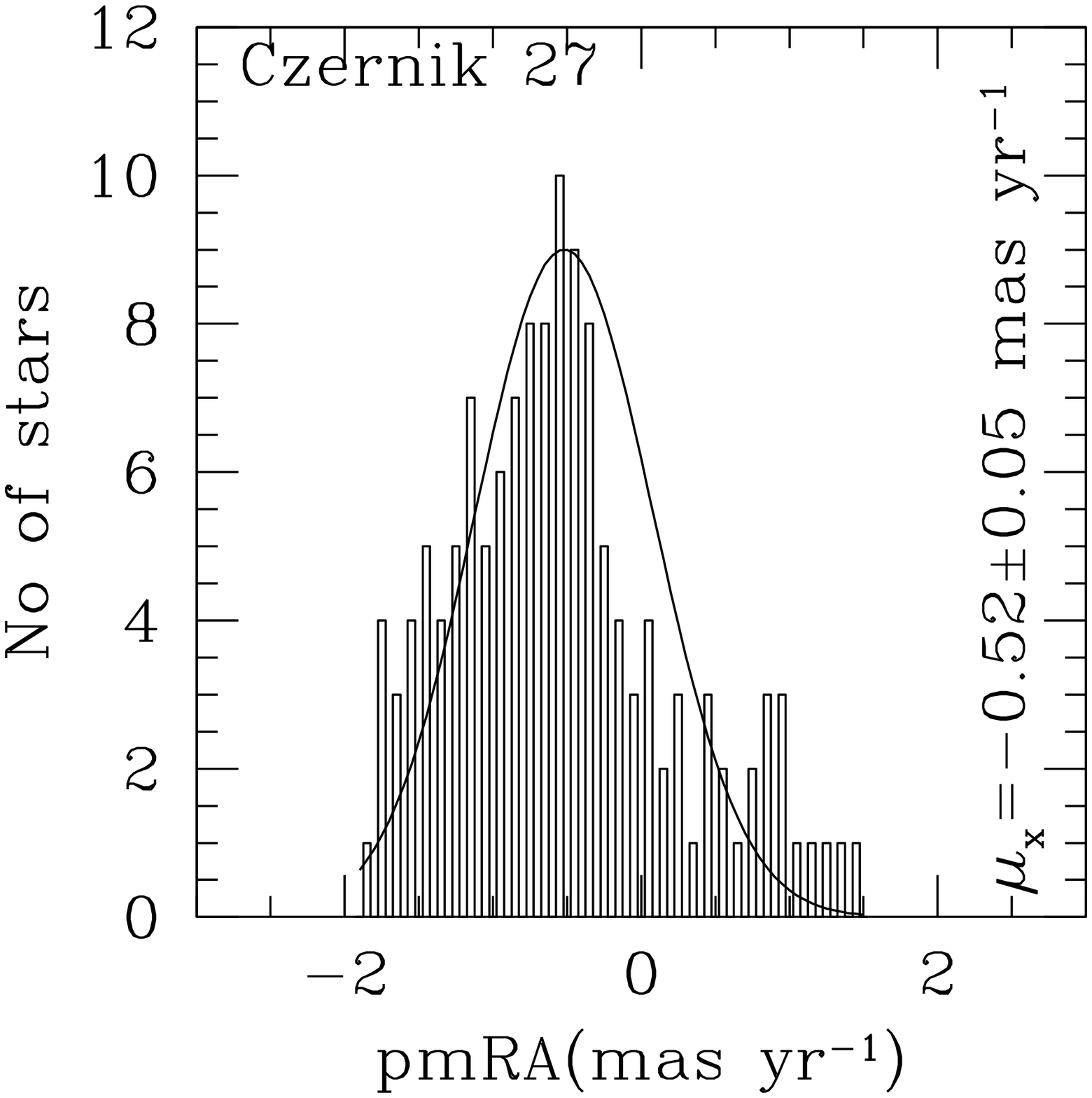}
\includegraphics[width=4.0cm,height=4.0cm]{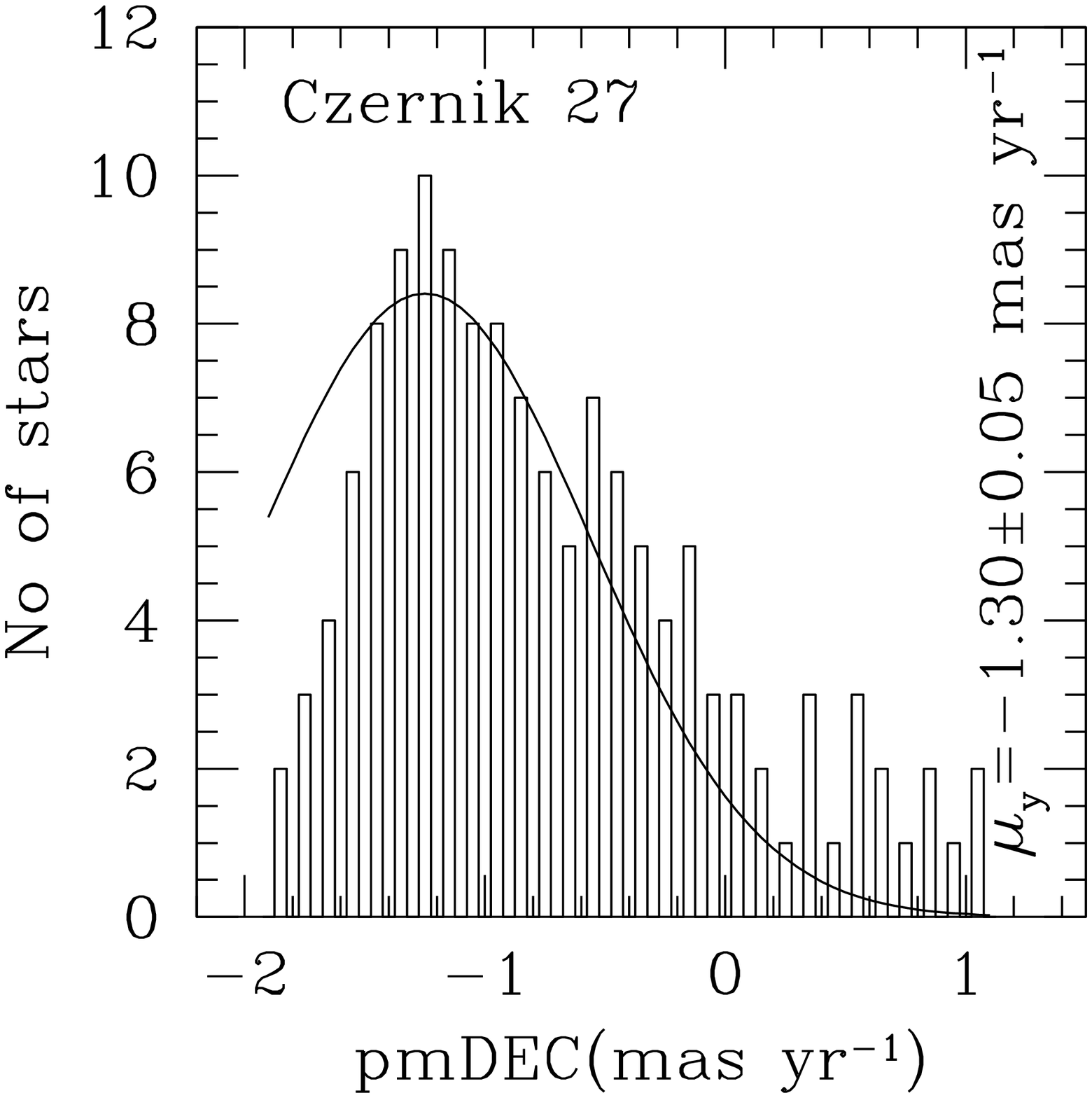}
}
\caption{Proper motion histograms of 0.1 mas/yr bins in right ascension and declination of the
candidate clusters. The Gaussian function fit to the central bins provides the mean values in
both directions as shown in each panel.}
\label{pm_hist}
\end{center}
\end{figure}

To determine the mean proper motion of the clusters we considered probable cluster members based on VPD and CMD for clusters Be 24 
and Cz 27 respectively as shown in Fig \ref{pm_cmd}. We have constructed the histograms and fitted the Gaussian function to the central
bin, which provides mean proper motion in both directions as shown in Fig.~\ref{pm_hist}. We have thus found the mean-proper motion of
Be 24 as $0.35\pm0.06$ mas yr$^{-1}$ and $1.20\pm0.08$ mas yr$^{-1}$ in RA and DEC directions, respectively while the same for cluster 
Cz 27 are found to be $-0.52\pm0.05$ mas yr$^{-1}$ and $-1.30\pm0.05$ mas yr$^{-1}$. The mean proper motion of the clusters is determined
as follows:\\

~~~~~~~~~~~~~~~~~~~~~~~~~~$\mu=\sqrt{(\mu_{x}^2+\mu_{y}^2)}$\\

The estimated value of mean proper motion is found to be $1.25\pm0.09$ mas yr$^{-1}$ and $1.40\pm0.05$ mas yr$^{-1}$
for the clusters Be 24 and Cz 27 respectively. Here the uncertainties are standard deviations.

\section{Structural parameters, colour-magnitude diagrams, reddening law and metallicity of the clusters} \label{sec:ana}

\subsection{Center estimation}

  \begin{figure}
    \centering
   \hbox{ 
   \includegraphics[width=4cm, height=4cm]{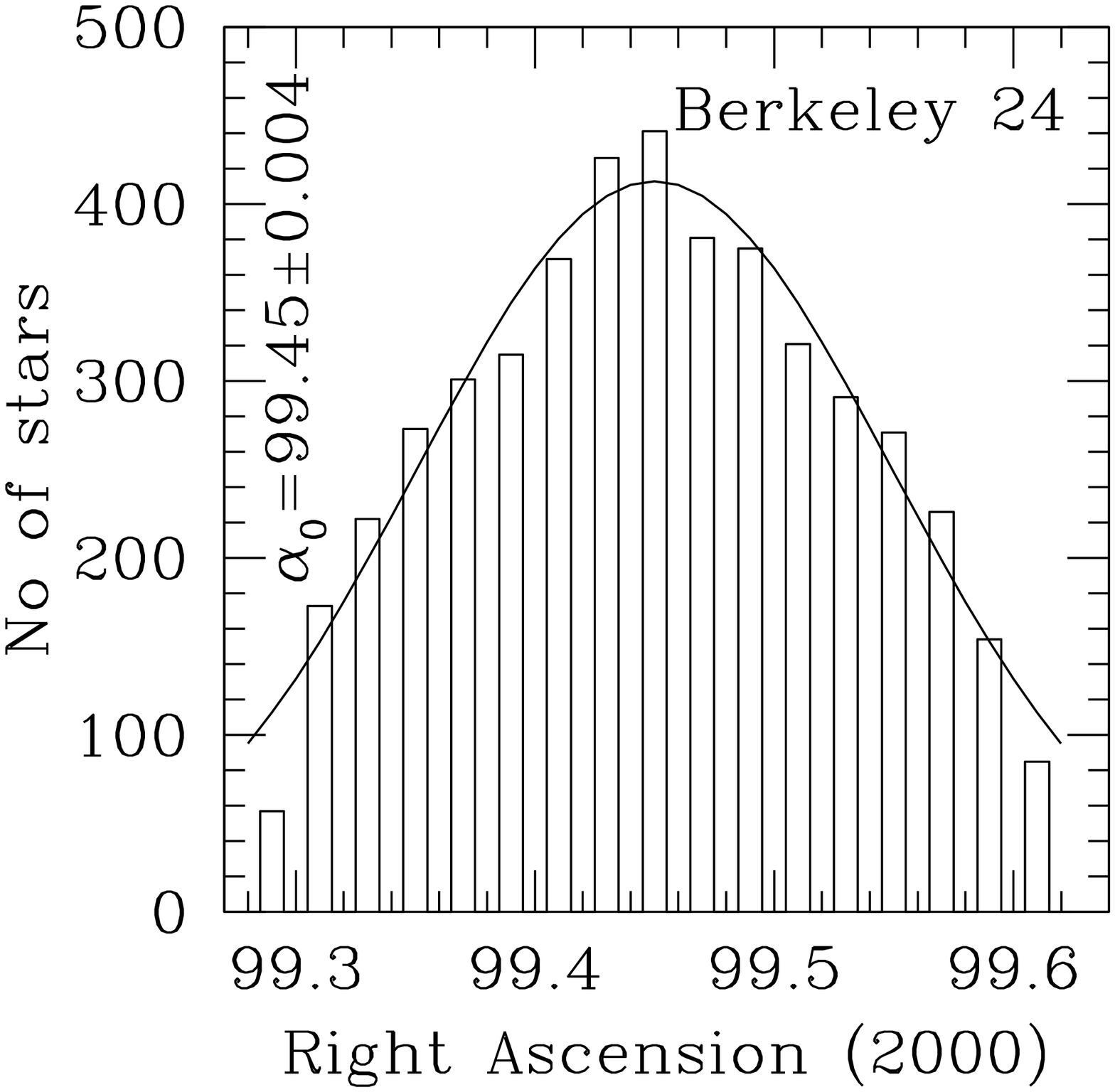}
    \includegraphics[width=4cm, height=4cm]{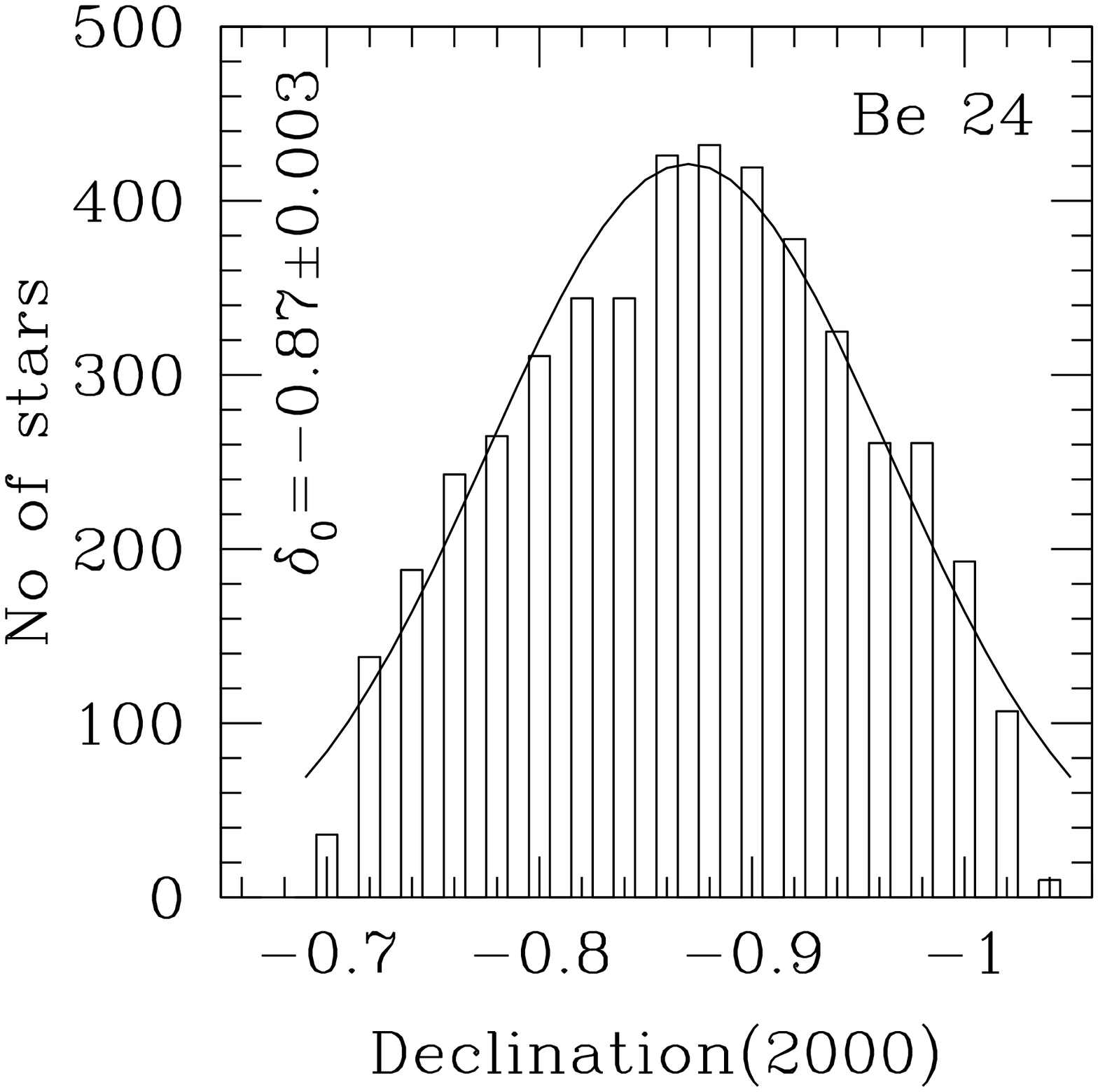}
}
   \hbox{ 
   \includegraphics[width=4cm, height=4cm]{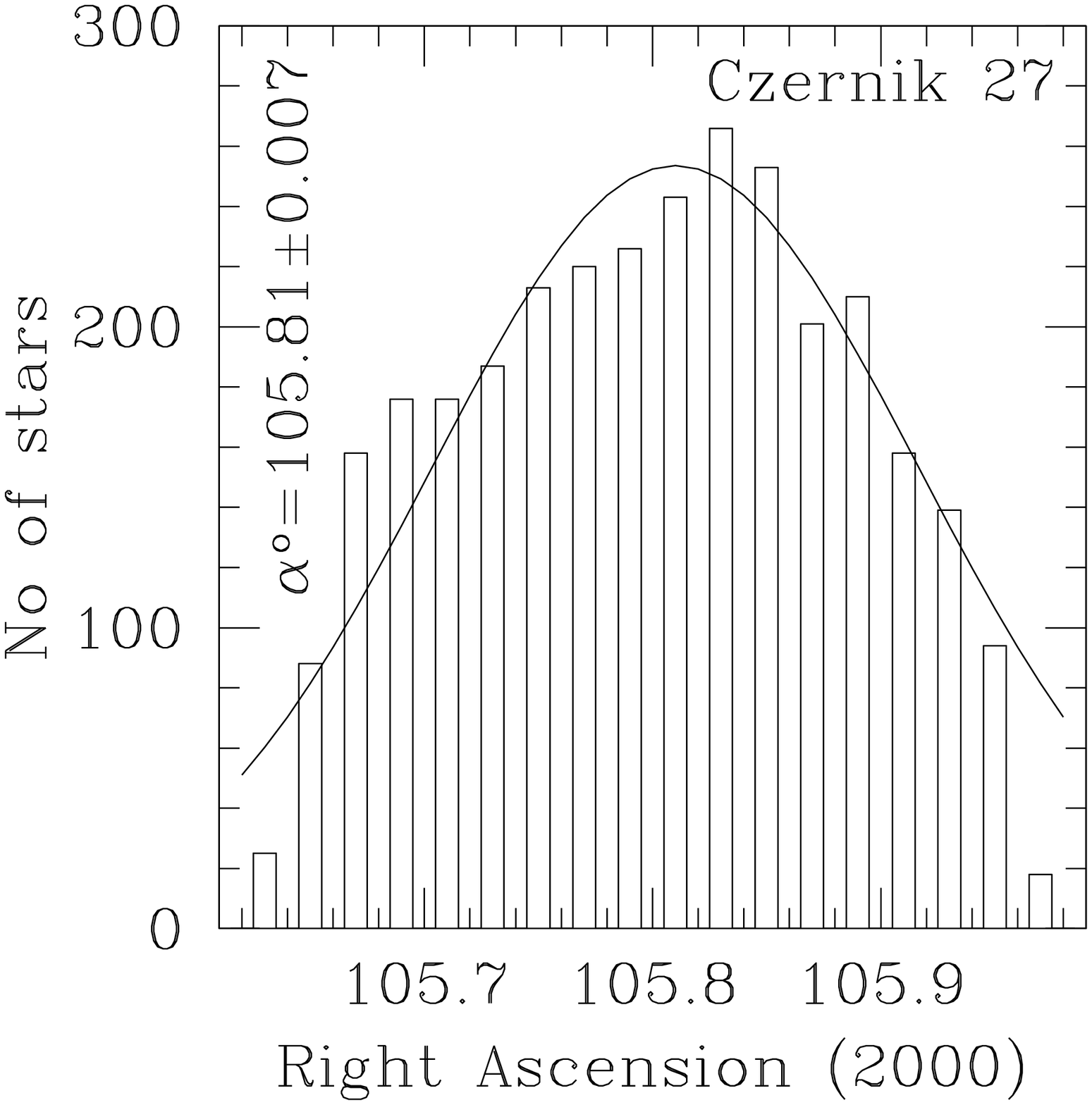}
    \includegraphics[width=4cm, height=4cm]{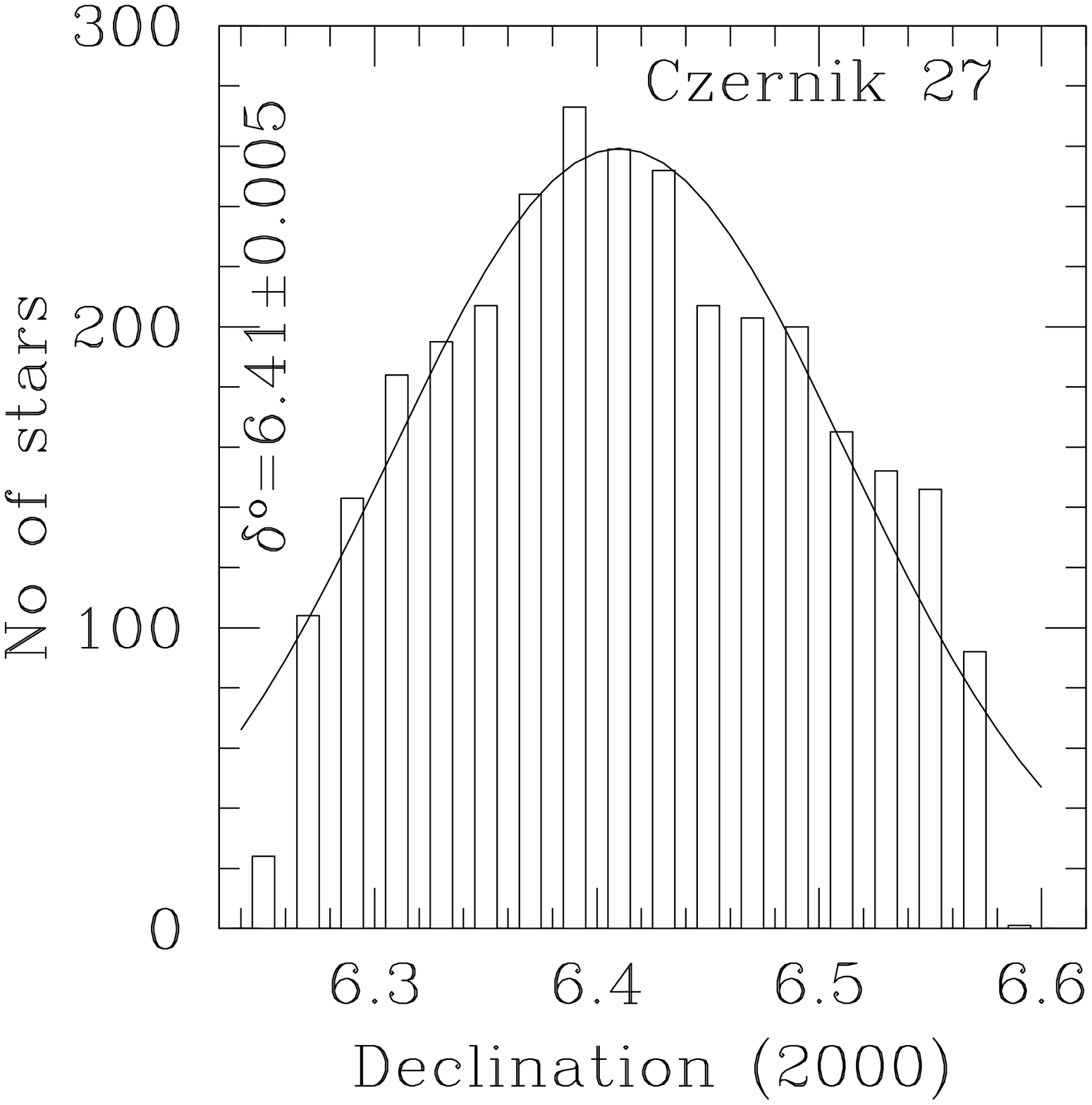}
}
\caption{Profiles of stellar counts across the clusters Be 24 (upper two panels) and Cz 27
(Lower two panels). Gaussian fits have been applied to derive the centroid of the two clusters in
right ascension and declination.}
  \label{center}
  \end{figure}

To study the shape of a star cluster, the first necessary step is to identify the cluster center. The location 
of maximum stellar density of the cluster's area is defined as the cluster centre. Earlier studies have
estimated cluster centres by visual inspection  (e.g., Becker \& Fenkart 1971, Romanishim \& Angel 1980),
but in the present analysis, we have applied the star-count method to the observed area of each cluster. To 
estimate the cluster center, we have plotted the histogram of star counts in Right Ascension (RA) and Declination
(DEC) of the stars using GDR2. For this purpose, the cluster area is divided into equal sized bin in RA and DEC.
The purpose of this counting process is to estimate the maximum central density of clusters. The cluster center is 
determined by fitting Gaussian profiles of star counts in RA and DEC, as shown in Fig.~\ref{center}. The best fit Gaussian
profiles provide the central coordinates of the clusters as $\alpha = 99.45\pm0.004$ deg and $\delta = -0.87\pm0.003$ deg
for Be 24 and $\alpha = 105.81\pm0.007$ deg and $\delta = 6.41\pm0.005$ deg for Cz 27. These fitted values of the cluster 
centers are in very good agreement with the values listed from the literature in Table~1. We have adopted these fitted centers
for further analysis.

\subsection{Size of the clusters}

\begin{figure}
\centering
\hbox{ 
\includegraphics[width=4cm, height=4cm]{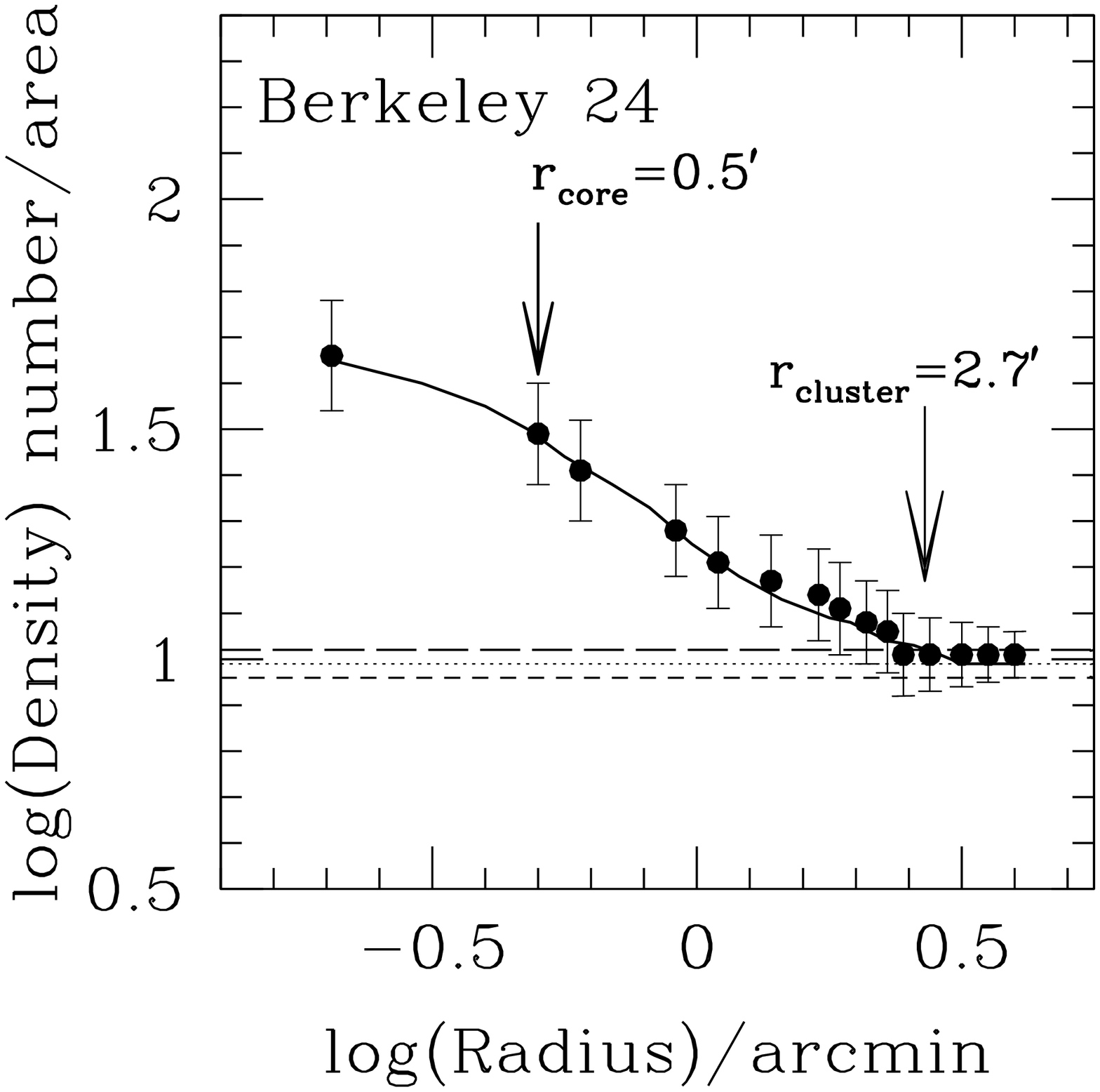}
\includegraphics[width=4cm, height=4cm]{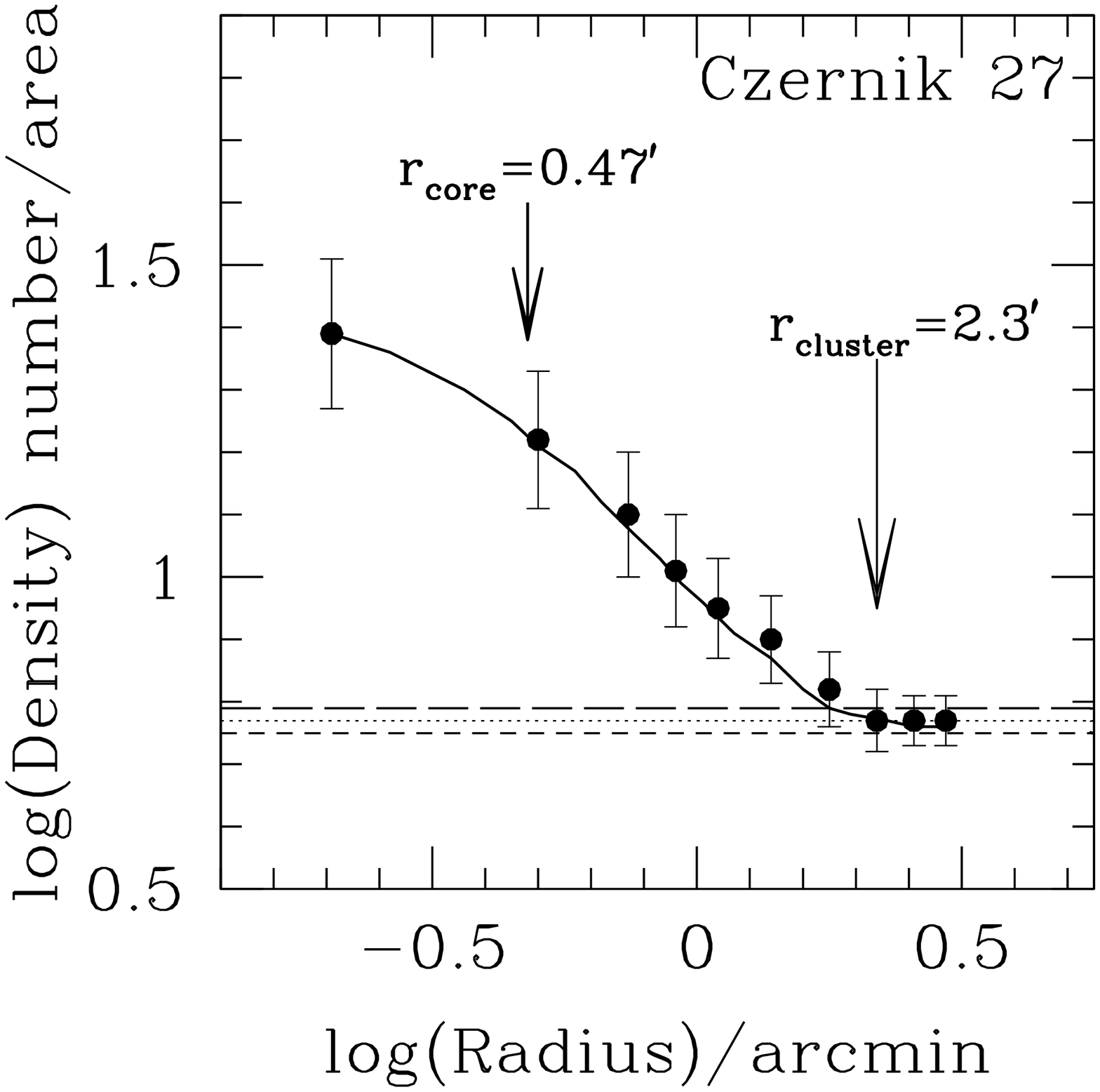}
}
\caption{Surface density distribution of the clusters Be 24 and Cz 27. 
Errors are determined from sampling statistics (=$\frac{1}{\sqrt{N}}$ where $N$ is 
the number of stars used in the density estimation at that point). The smooth line represent 
the fitted profile whereas dotted line shows the background density level. Long and short dashed
lines represent the errors in background density. Arrow indicates the location of cluster 
and core radii.}
\label{dens}
\end{figure}

Determination of the cluster extent is necessary for reliable estimation of all relevant fundamental parameters of open
clusters. For this purpose, we have derived the surface stellar density by performing star counts in concentric
rings around our fitted centres of the clusters, and then divided by their respective areas. In Fig.~\ref{dens} we
have shown the density profile (log(radius) versus log(density)) for both the clusters. This density distribution shows 
a peak near the cluster center and becomes consistent with being flat after a certain point where the cluster density 
merges into the field star density. Therefore, we have considered 2.7(log(radius)=0.43) and 2.3 (log(radius)=0.36) arcmin
as the cluster radius for Be 24 and Cz 27, corresponding to 3.4 and 3.2 pc, respectively. A smooth continuous line 
representing a King (1962) profile:\\

~~~~~~~~~~~~~~~~~~~~$f(r) = f_{bg}+\frac{f_0}{1+(r/r_{c})^2}$\\

has been fitted to the radial density profiles. Here $r_{c}$ , $f_{0}$  and $f_{bg}$ are the core radius, central 
density, and the background density respectively. The core radius is defined as a distance where the
stellar density becomes half of the central density. By fitting the King model to the radial density profile, 
we found core radii as 0.5 (log(core radius)=-0.30) and 0.47 (log(core radius)=-0.32) arcmin as shown in Fig.~\ref{dens}
for Be 24 and Cz 27 respectively. The estimated values of core radii along with the other structural parameters are listed in 
Table~\ref{stru_para} for both the clusters. The location of cluster radii and core radii indicated by arrow 
in the radial density profiles. The cluster limiting radius, $r_{lim}$, was calculated by comparing $f(r)$ to a 
background density level, $f_{b}$, defined as\\

\begin{table}
\centering
\caption{Structural parameters of the clusters Be 24 and Cz 27. Background and central 
density are in the unit of stars per arcmin$^{2}$. Core radius ($r_c$) and tidal radius ($R_t$) are 
in arcmin and pc.
}
\vspace{0.5cm}
\begin{center}
\small
\begin{tabular}{cccccccc}
\hline\hline
Name & $f_{0}$ &$f_{b}$& $r_{c}$&$r_{c}$&$R_{t}$& $R_{t}$ & $\delta_{c} $ \\
&&& arcmin & parsec & arcmin & parsec
\\
Be 24 & $52.3$&$9.08$&$0.5$&$0.64$&$5.4$&$6.7$&$6.7$ \\
Cz 27  & $29.1$&$5.81$&$0.47$&$0.76$&$3.8$&$6.3$&$6.0$ \\
\hline
\end{tabular}
\label{stru_para}
\end{center}
\end{table}
~~~~~~~~~~~~~~~~~~$f_{b}=f_{bg}+3\sigma_{bg}$\\

where $\sigma_{bg}$ is uncertainty of $f_{bg}$. Therefore, $r_{lim}$ was calculated according to 
the following formula\\

~~~~~~~~~~~~~~~$r_{lim}=r_{c}\sqrt(\frac{f_{0}}{3\sigma_{bg}}-1)$\\

The value for the limiting radius was found to be 3.0 and 3.8 arcmin for Be 24 and Cz 27, respectively.
$r_{c}$ and $r_{lim}$ can be combined to calculate the concentration parameter
$c = log (\frac{r_{lim}} {r_{c}})$ (Peterson \& King, 1975) to further characterize the structure of clusters
in the Milky Way Galaxy. In the present study, we found the concentration parameters to be 0.7 for both 
clusters. Maciejewski \& Niedzielski (2007) reported that $r_{lim}$ may vary for individual clusters 
from 2$r_{c}$ to 7$r_{c}$. In this case, both clusters show good agreement with Maciejewski \& Niedzielski (2007). 

The density contrast parameter is estimated for clusters using the following relationship\\

~~~~~~~~~~~~~~~~~~$\delta_{c} = 1 +\frac{f_{0}}{f_{bg}}$\\

We found values of $\delta_{c}$ of 6.7 and 6.0 for Be 24 and Cz 27, respectively. These values of $\delta_{c}$ are
lower than the values ($7\le \delta_{c}\le 23$) derived for compact star clusters by Bonatto \& Bica (2009). This 
implies that both clusters studied here are sparse clusters.

The tidal radius of open clusters is the distance from the cluster core at which the gravitational influence of 
the Galaxy is equal to that of the open cluster core. The tidal radius of a cluster can be estimated using the
following procedure-

The Galactic mass $M_{G}$ inside a Galactocentric radius $R_{G}$ is given by (Genzel \& Townes, 1987),\\

~~~~~~~~~~~~~~~~$M_{G}=2\times10^{8} M_{\odot} (\frac{R_{G}} {30 pc})^{1.2}$\\

The value of $M_{G}$ is found to be $2.8 \times 10^{11}$ $M{\odot}$ and $3.2 \times 10^{11}$ $M{\odot}$ for clusters Be 24
and Cz 27 respectively.

Using the formula by Kim et al. (2000), tidal radius $R_{t}$ of a cluster is, \\

~~~~~~~~~~~~~~~~~$R_{t}=(\frac{M_{c}} {2M_{G}})^{1/3}\times R_{G}$\\

where $R_{t}$ and  $M_{c}$ are the tidal radii and total mass (see Sect.~8) of the clusters, respectively.
We derive values of the tidal radius of 6.7 and 6.3 pc for Be 24 and Cz 27, respectively. These are also listed in
Table~\ref{stru_para}. Our measurement is reliable because it is based on the probable members of the clusters 
selected from VPD.

\subsection{Reddening law}

The plots of two-colour diagrams (TCDs) for various sets of two colours are very useful tools to 
estimate interstellar reddening and to understand the properties of the extinction law in the direction of the clusters.

\subsubsection{Total-to-selective extinction value}

\begin{figure}
    \centering
   \hbox{ 
    \includegraphics[width=4cm, height=5cm]{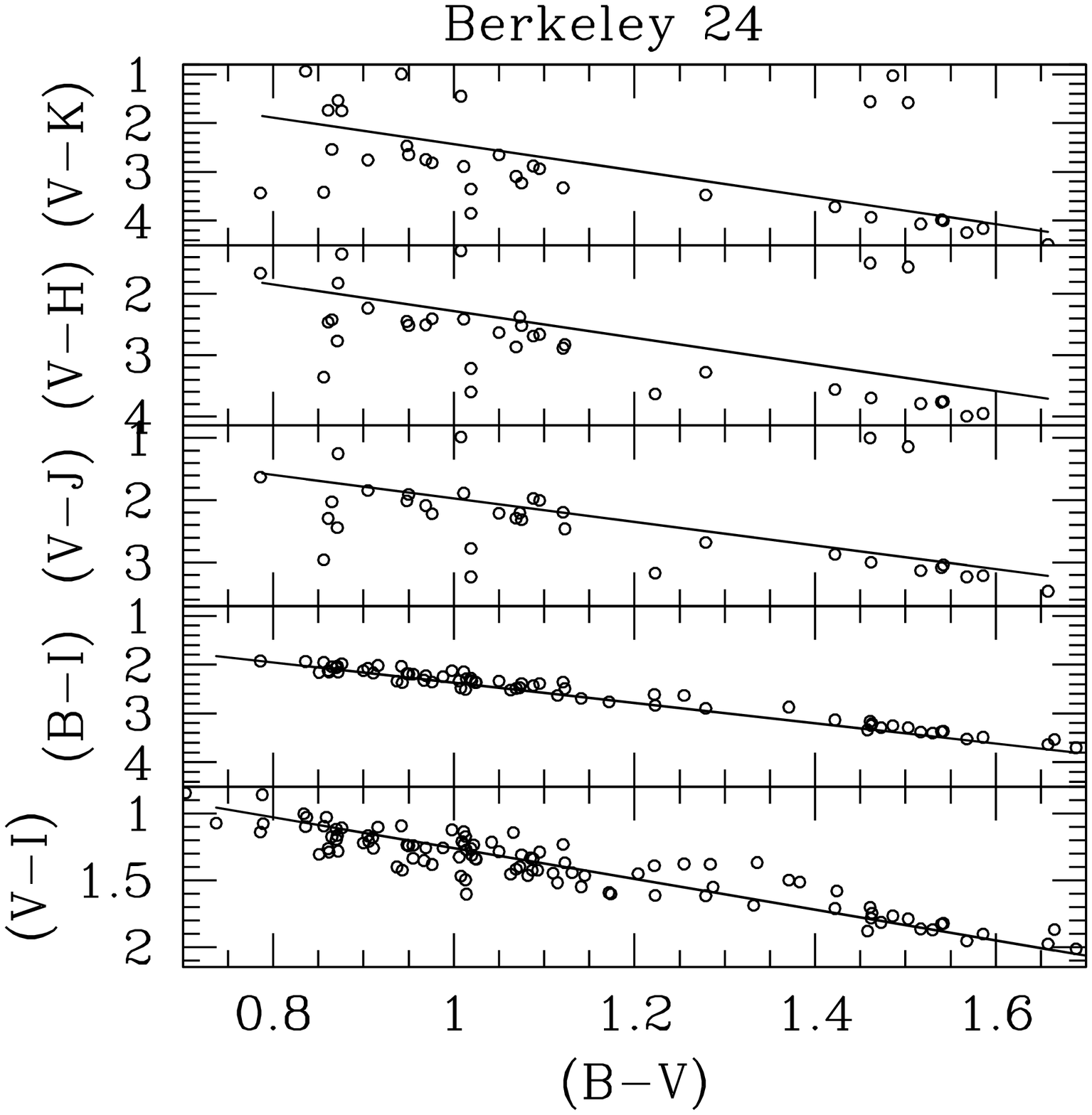}
   \includegraphics[width=4cm, height=5cm]{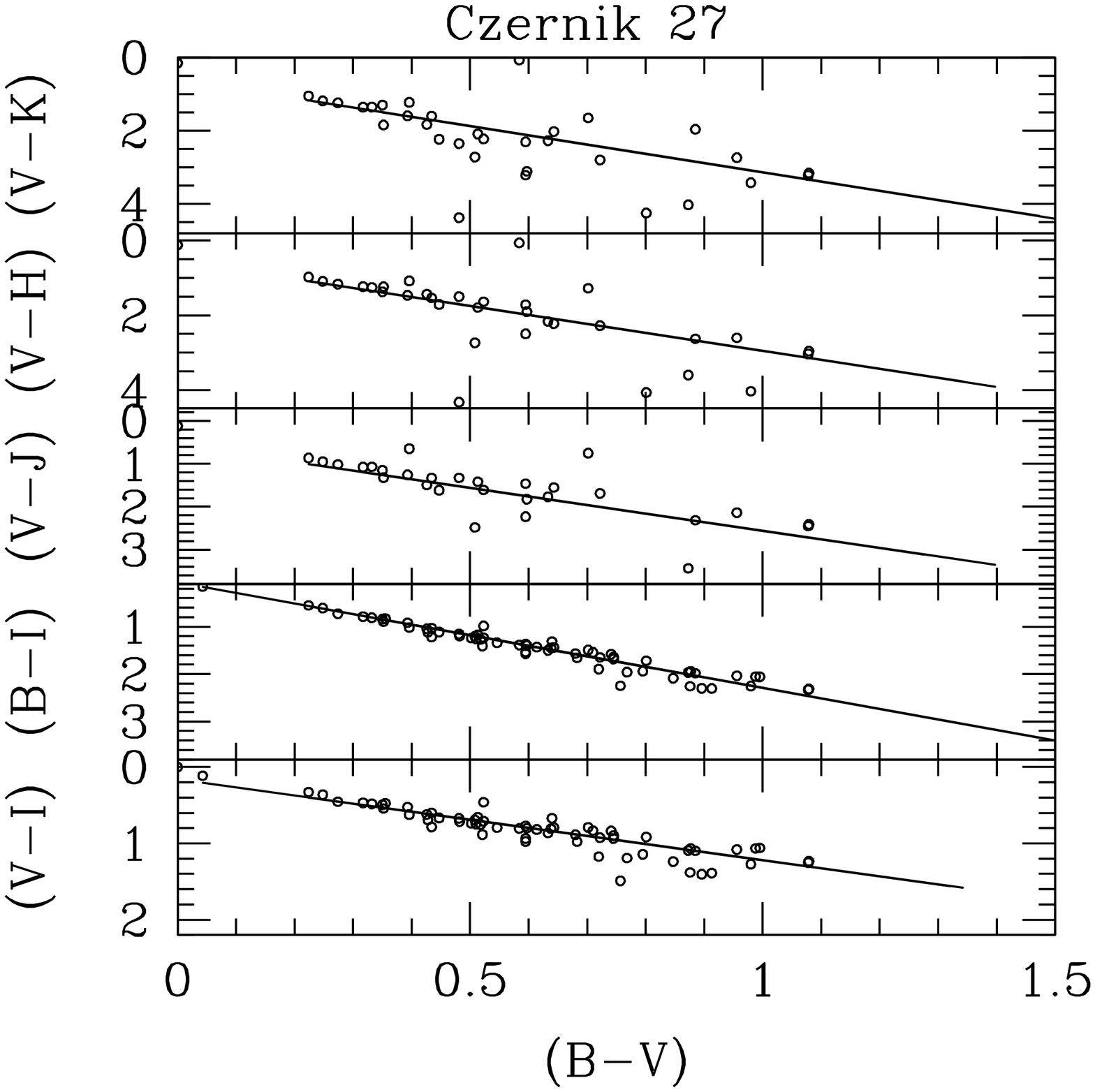}
}
\caption{The $(\lambda-V)/(B-V)$ TCD for the stars within cluster extent of clusters Be 24 and  Cz 27. The continuous lines
represent the slope determined through least-squares linear fit.}
  \label{cc2}
  \end{figure}

Reddening is an important basic parameter of a star cluster, since it can
significantly affect the determination of other fundamental parameters. To
derive the characteristics of the extinction law, it is important to analyse
two colour diagrams.  The emitted photons of cluster stars are scattered and
absorbed in the interstellar medium by dust particles, which leads to deviation
of colours from their intrinsic values. The normal Galactic reddening law is
often not applicable in the line of sight to clusters (e.g., Sneden et al. 1978).
Chini \& Wargue (1990) suggested $(V-\lambda)/(B-V)$ TCDs to examine the nature
of the reddening law. Here, $\lambda$ denotes nearly any filter other than $V$.
We have studied the reddening law for both clusters by drawing TCDs as shown in
Fig.~\ref{cc2}. Since, the stellar colour values are found to be linearly
dependent on each other, then a linear equation is applied to calculate the
slope $(m_{cluster})$ of each TCD. We have estimated total to selective
extinction using the relation given by Neckel \& Chini (1981):\\

$R_{cluster}=\frac{m_{cluster}}{m_{normal}}\times R_{normal}$\\

where $m_{cluster}$ is the typical value of the slope in a TCD and $R_{normal}$ (numerically 3.1) 
is the normal
value of total to selective extinction ratio. We have estimated $R_{cluster}$ in different passbands
to be $3.1\le R_{cluster}\le3.5$ which is slightly larger than the normal value. Thus, reddening law is found to
be normal towards the cluster region for both clusters.

  \begin{figure}
    \centering
    \includegraphics[width=8cm, height=8cm]{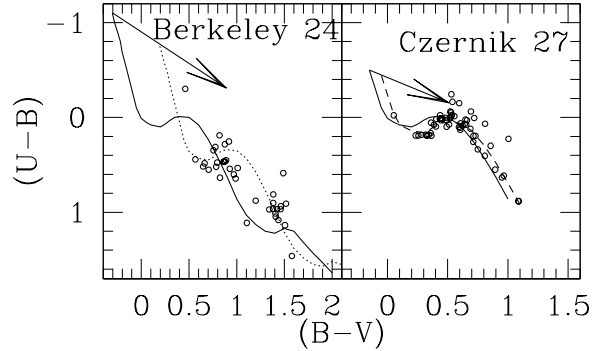}
\vspace{-1cm}
\caption{The $(U-B)$ versus $(B-V)$ colour-colour diagram of the clusters.
The continuous curve represents locus of Schmidt-Kaler's (1982) ZAMS for
solar metallicity. Dotted and dashed lines are the shifted Schmidt-Kaler ZAMS with the values 
given in the text.}
  \label{cc}
  \end{figure}

\subsubsection{$(U-B)$ versus $(B-V)$ diagram}

A knowledge of reddening is very important for the intrinsic properties of cluster stars. In the absence 
of spectroscopic data, the $(B-V), (U-B)$ colour-colour diagram is widely used for the reddening estimation 
(e.g., Becker \& Stock 1954).
To estimate the reddening towards the cluster region, we have plotted $(U-B)$ versus $(B-V)$ diagram for both the
clusters as shown in Fig.~\ref{cc} using stars within cluster extent. The intrinsic zero-age main-sequence (ZAMS) given by 
Schmid-kaler (1982) is fitted by the continuous curve assuming the slope of reddening $E(U-B)/E(B-V)$ as 0.72. By fitting ZAMS 
to the MS, we have calculated mean value of $E(B-V)=0.45\pm0.05$ mag for Be 24 and $E(B-V)=0.15\pm0.05$ for Cz 27,
respectively. 
Our derived values of reddening agree fairly well with the values estimated by others as discussed in Sect.~1.

\subsubsection{Interstellar extinction in near-IR} \label{sec:extir}

   \begin{figure}
    \centering
   \includegraphics[width=8cm, height=8cm]{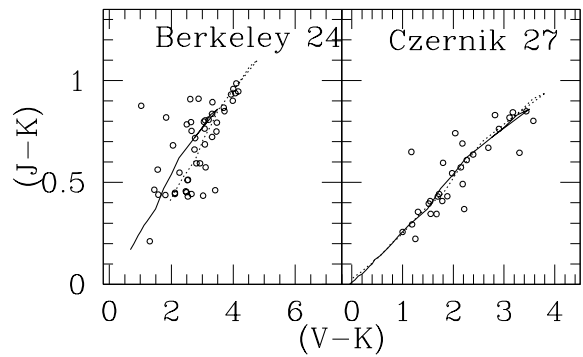}
\vspace{-1.2cm}
\caption{The plot of $(J-K)$  versus $(V-K)$ colour-colour diagram of the clusters for
the stars within the cluster radius. Solid line is the ZAMS taken from
Caldwell et al. (1993)}
  \label{cc_jk}
  \end{figure}

$2MASS$ $JHKs$ data combined with optical data is used to study the interstellar extinction law towards
the cluster region. The $K_s$ magnitude is converted into $K$ magnitude using the formulations by 
Persson et al. (1998). The $(J-K)$ versus $(V-K)$ diagram for both the clusters are shown in Fig.~\ref{cc_jk}. The ZAMS is 
taken from Caldwell et al. (1993) for $Z=$ 0.01 is shown by a solid line. The fit of ZAMS provides $E(J-K) = 0.23\pm0.03$ mag
and $E(V-K) = 1.23\pm0.02$ mag for Be 24 and $E(J-K) = 0.06\pm0.02$ mag and $E(V-K) = 0.33\pm0.01$ mag for Cz 27. The 
ratios $\frac{E(J-K)}{E(V-K)} \sim 0.18\pm0.06$ for Be 24 and Cz 27 are in good agreement with the normal interstellar 
extinction value of 0.19 given by Cardelli (1989). The scatter is primarily due to large error in the $JHK$ data. 

\subsubsection{$(B-V)$ versus $(J-K)$ diagram}

\begin{figure}
    \centering
   \includegraphics[width=8cm, height=8cm]{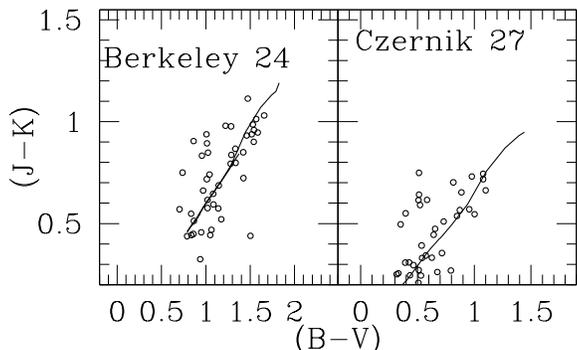}
\vspace{-1.8cm}
\caption{The plot of $(B-V)$  versus $(J-K)$ colour-colour diagram of the clusters for
the stars within the cluster radius. Solid line is the theoretical isochrone of Girardi et al. (2000)}
  \label{cc_bvjk}
  \end{figure}

We have plotted $(B-V)$ versus $(J-K)$ colour-colour diagram for clusters Be 24 and Cz 27, as shown in Fig.~\ref{cc_bvjk}.
To know the relationship between these two colours, we have used the theoretical isochrones given by Girardi (2000). The 
colour excess $E(B-V)$ and $E(J-K)$ for Be 24 is found to be 0.45 and 0.23 mag respectively, whereas for Cz 27 we find 0.15
and 0.08 mag. We got the ratio, $\frac{E(J-K)}{E(B-V)}\sim0.53$ for both the clusters. In the literature, the above value 
is mentioned as 0.56, which is computed by using the following relations, $A_{k}=0.618\times E(J-K)$ (Mathis 1990); 
$A_{k}=0.122\times A_{v}$ (Cardeli, Clayton \& Mathis 1989) and $A_{v}=3.1\times E(B-V)$. Our estimated
value is in good agreement with the literature.

\subsubsection{$(J-H)$ versus $(J-K)$ diagram}

\begin{figure}
    \centering
   \includegraphics[width=8cm, height=8cm]{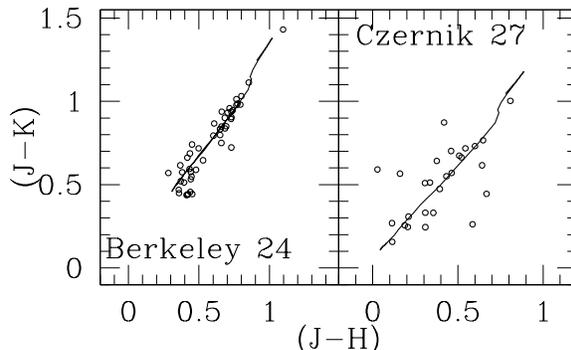}
\vspace{-1.8cm}
\caption{The plot of $(J-H)$  versus $(J-K)$ colour-colour diagram of the clusters for
the stars within the cluster radius. Solid line is the theoretical isochrone of Girardi et al. (2000)}
  \label{cc_jhjk}
  \end{figure}

$(J-H)$ versus $(J-K)$ colour-colour diagram for clusters Be 24 and Cz 27 is shown in Fig.~\ref{cc_jhjk}.
Stars plotted in this figure are within the cluster extent. The isochrone shown by the solid line is taken from 
Girardi et al. (2000). From this figure we have found $E(J-H)=0.14\pm0.05$ and $E(J-K)=0.24\pm0.02$ for Be 24 and for Cz 27, 
the above values are $0.05\pm0.03$ and $0.10\pm0.05$. The ratio $\frac{E(J-H)}{E(J-K)}=0.58\pm0.06$ for both the clusters 
are in good 
agreement with the normal interstellar extinction value 0.55 as suggested by Cardeli et al. (1989). Scattering is larger due
to large errors in $J$, $H$ and $K$ magnitudes. We can estimate the reddening, $E(B-V)$ using the following relations:\\

~~~~~~$E(J-H)=0.309\times E(B-V)$\\

~~~~~~$E(J-K)=0.48\times E(B-V)$\\

Our estimated values of reddening $E(B-V)=0.45$ and 0.16 for the cluster Be 24 and Cz 27, using the $2MASS$ colours are 
similar to those obtained from the $(U-B)$ versus $(B-V)$ colour-colour diagram. Thus it validates the use of the $2MASS$
colours for the $E(B-V)$ estimation when only $2MASS$ data are available- for e.g. in highly extincted regions.

\subsection{Metallicity of the clusters derived from photometry}

The metallicity of stars is an important tool to explore the chemical structure and evolution of
the Galaxy. The metallicities of the clusters Be 24 and Cz 27 have not been estimated in previous 
studies. For the estimation of metallicity, we have adopted a method, which is discussed in Karaali et al. (2003, 2005, 2011) 
using $UBV$ data. The procedure in this method is based on $F-G$ spectral type main-sequence stars of the cluster. Thus, we 
selected 69 stars for Cz 27 and 12 stars for Be 24 with colour range $0.3\leq  (B-V)_0\leq  0.6$ mag consistent with
the colours of the F-G spectral types.

  \begin{figure}
    \centering
   \hbox{ 
   \includegraphics[width=4cm, height=4cm]{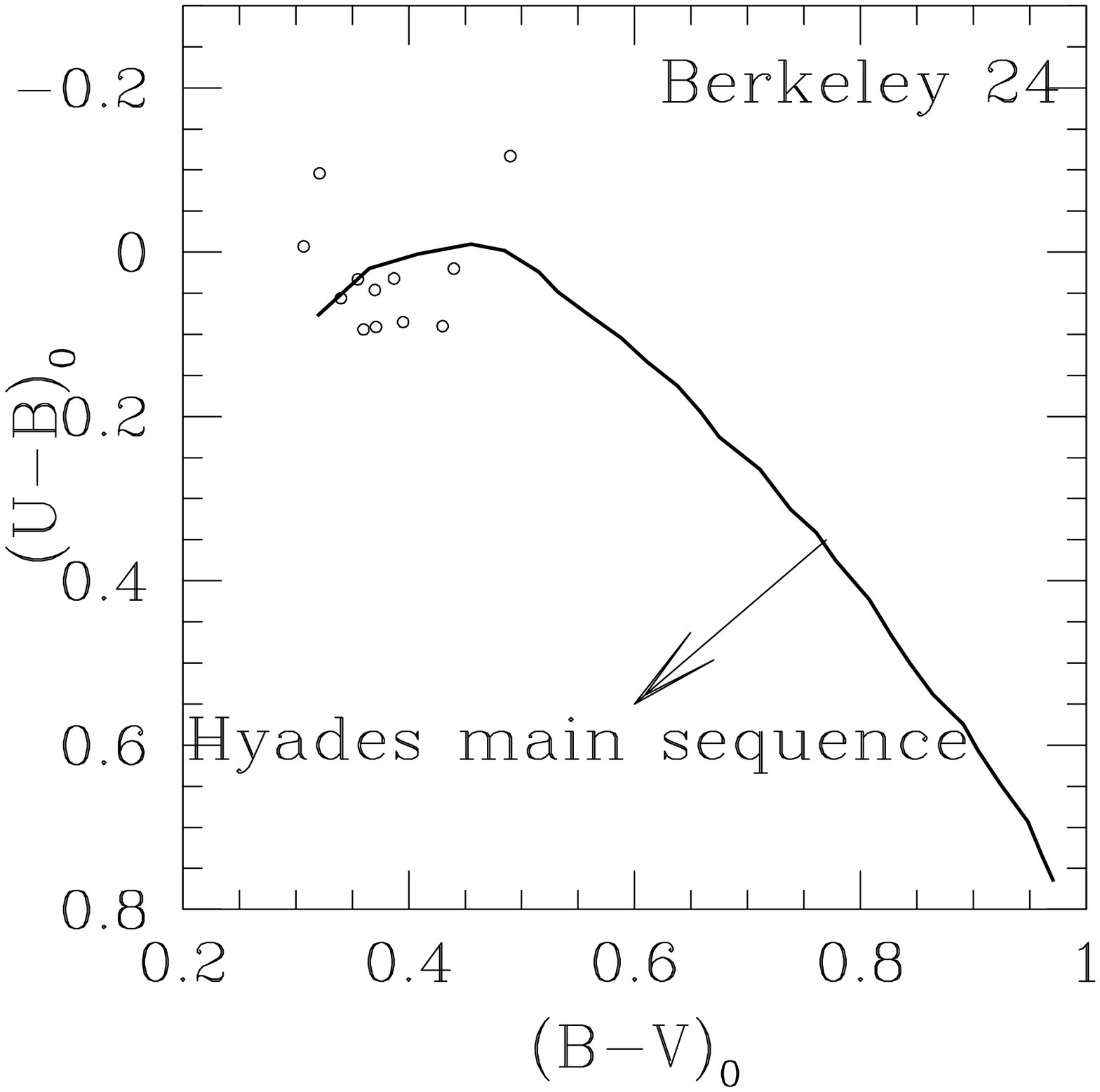}
    \includegraphics[width=4cm, height=4cm]{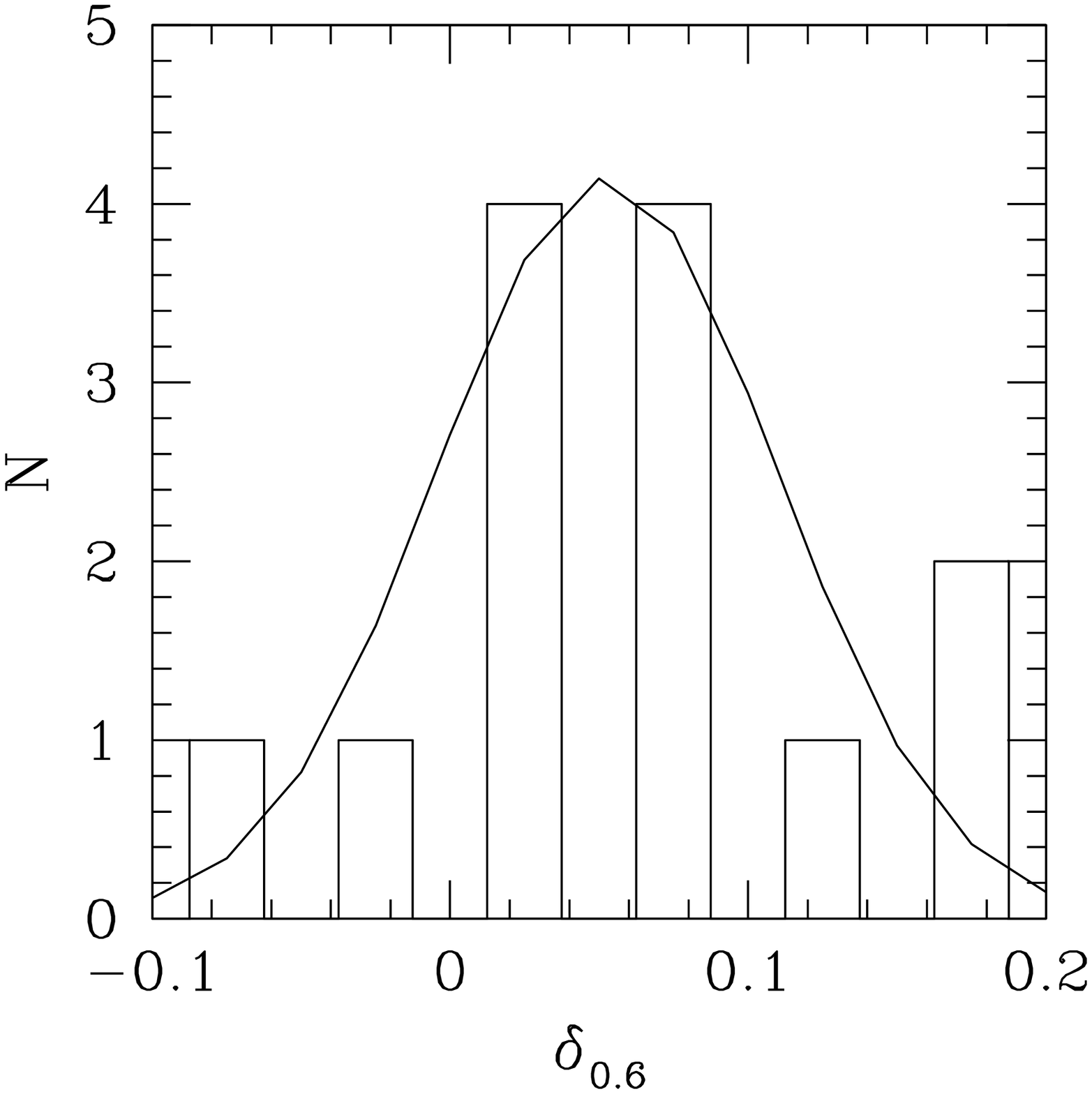}
}
   \hbox{ 
   \includegraphics[width=4cm, height=4cm]{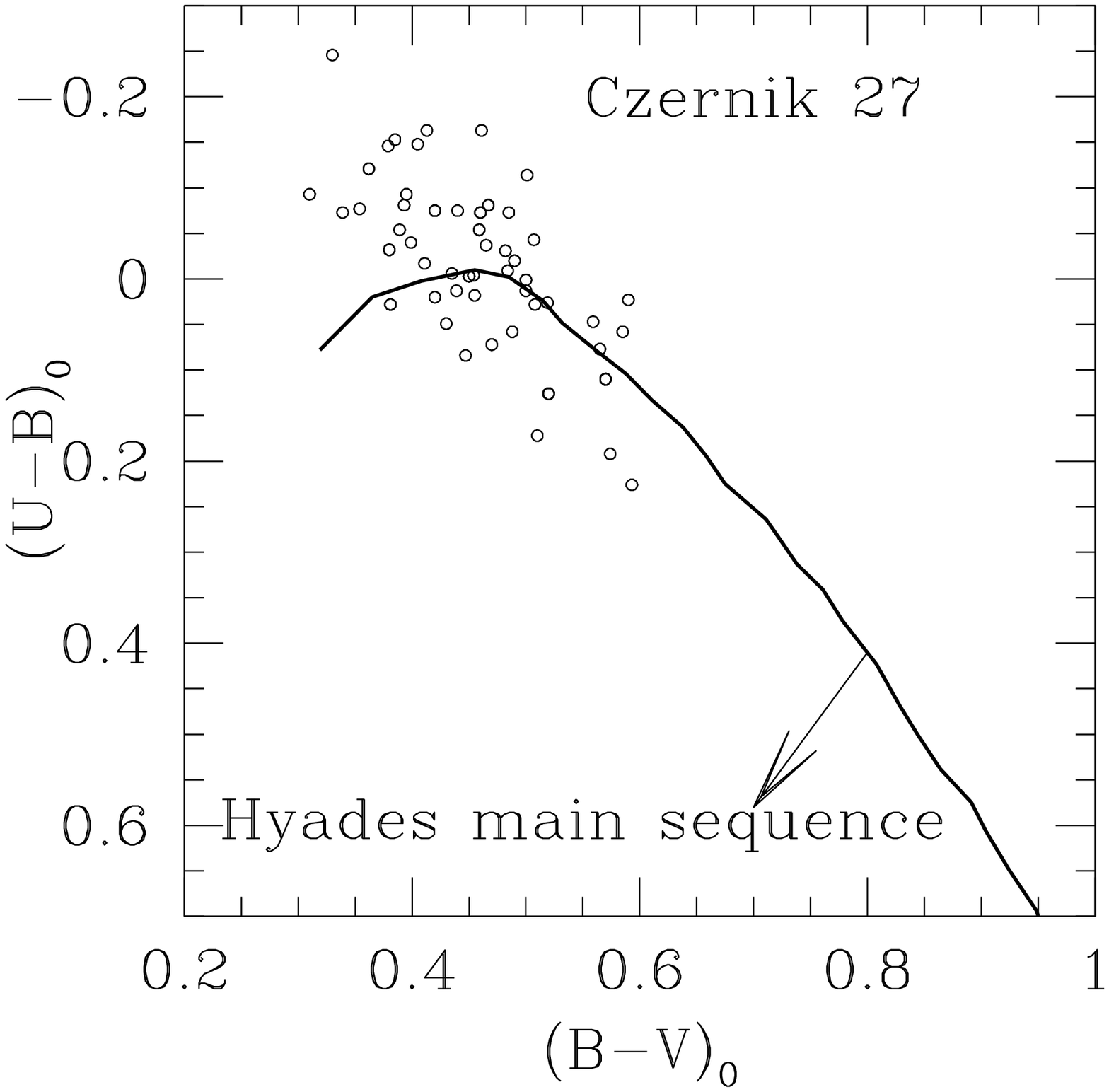}
    \includegraphics[width=4cm, height=4cm]{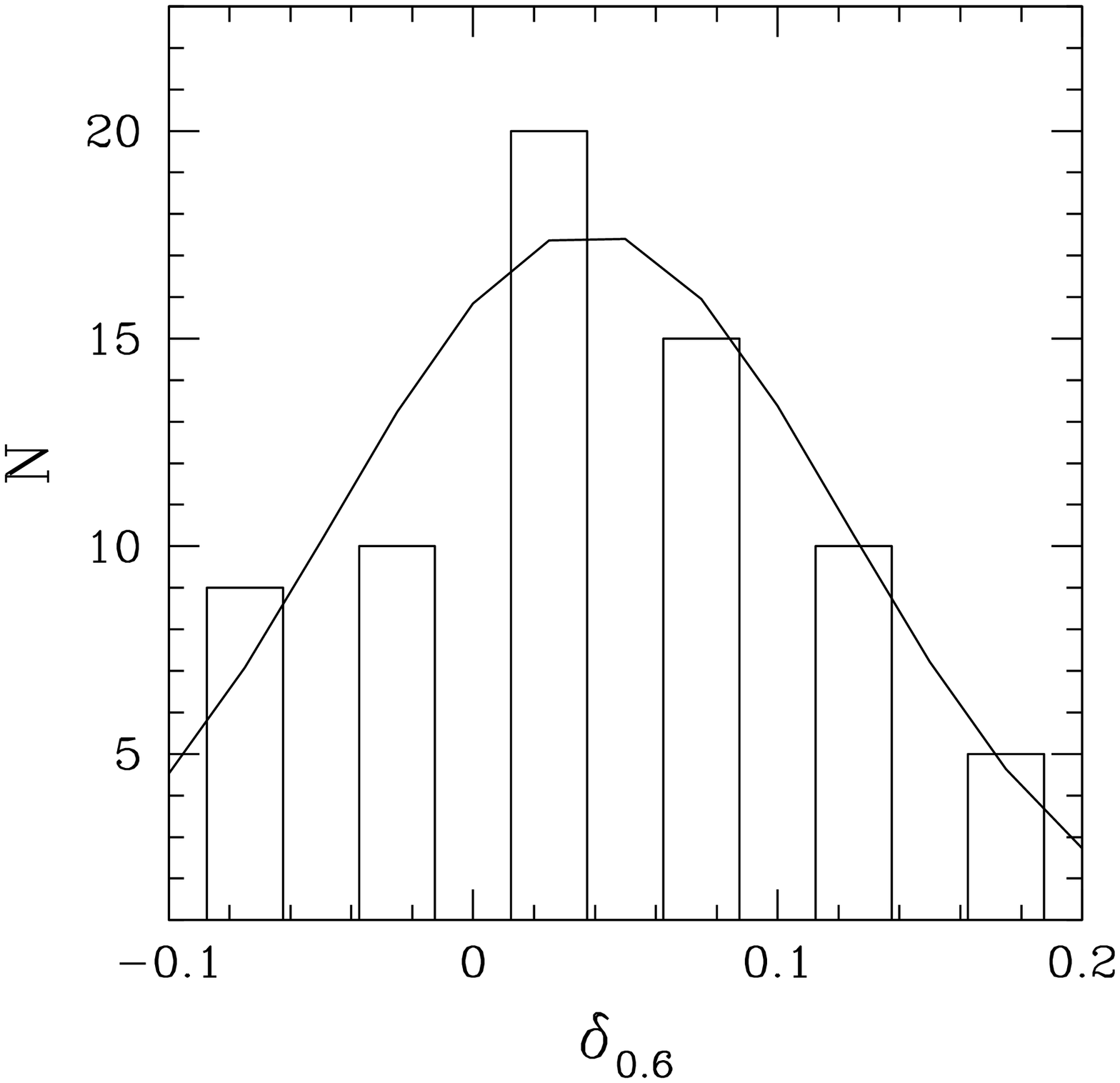}
}
\caption{The $(U-B)_{0} vs (B-V)_{0}$ two colour diagrams and histograms for the normalized UV-excess
for 12 (Be 24) and  69 (Cz 27)  main sequence stars used for the metallicity estimation of Be 24 and 
Cz 27. The solid line in the two colour diagram is Hyades main sequence and solid line in histogram is 
the Gaussian fit.}
  \label{met}
  \end{figure}
We have estimated the normalized ultraviolet (UV) excess for the selected stars, which is defined as
the differences  between a stars' de-reddened $(U-B)_{0}$  colour indices and the one corresponding
to the  members of the Hyades  cluster with the  same de-reddened  $(B-V)_{0}$  colour index, that is 
$\delta=(U-B)_{0,H}-(U-B)_{0,S}$. Here the subscripts $H$ and $S$ indicate Hyades and stars, respectively. 
In order to  utilize the method described  in Karaali et al.(2011), we have determined normalized $UV$ excesses of the 
selected stars as described above and normalized their $\delta$ differences to the UV-excess at $(B-V)_{0}=0.6$ mag, 
i.e. $\delta_{0.6}$. $(U-B)_0$ vs $(B-V)_0$ colour-colour diagram and corresponding histograms of the normalized 
$UV$ excesses ($\delta_{0.6}$) of the 
selected stars for Be 24 and Cz 27  are presented in Fig.~\ref{met}. By  fitting a Gaussian function to this histogram, 
we have calculated the normalized UV  excess of the cluster  as $\delta_{0.6}=0.051\pm 0.002$ mag for Be 24
and $\delta_{0.6}=0.039\pm 0.002$ mag for Cz 27.  Here, the uncertainty is given as the statistical uncertainty of
the peak  of the Gaussian. Then,  we have determined the  metallicity ($[Fe/H]$) of  the clusters by  evaluating this 
Gaussian peak value in the equation discussed in Karaali et al. (2011):\\

  \begin{figure*}
    \centering
   \hbox{ 
   \includegraphics[width=9cm, height=8cm]{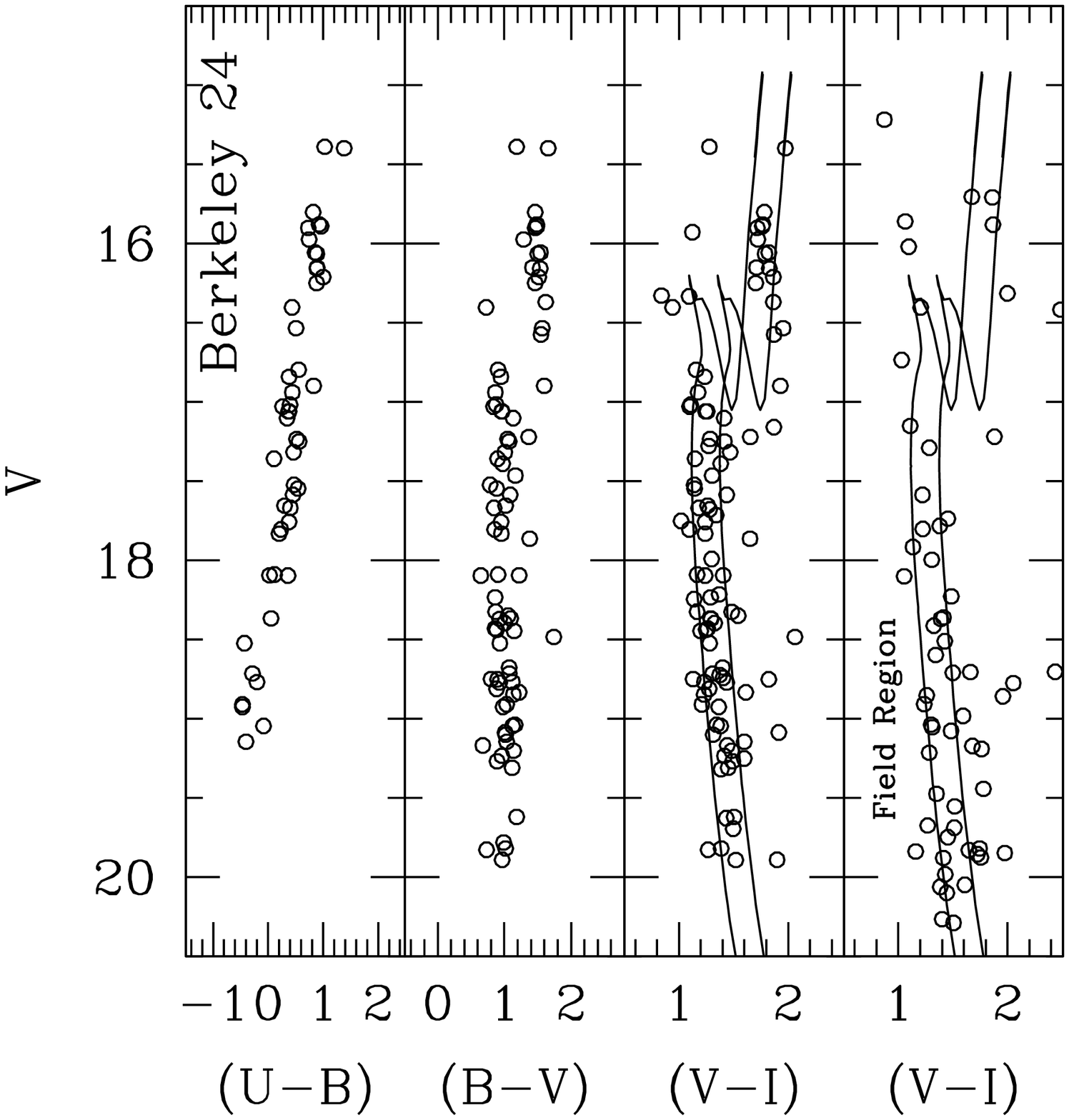}
    \includegraphics[width=9cm, height=8cm]{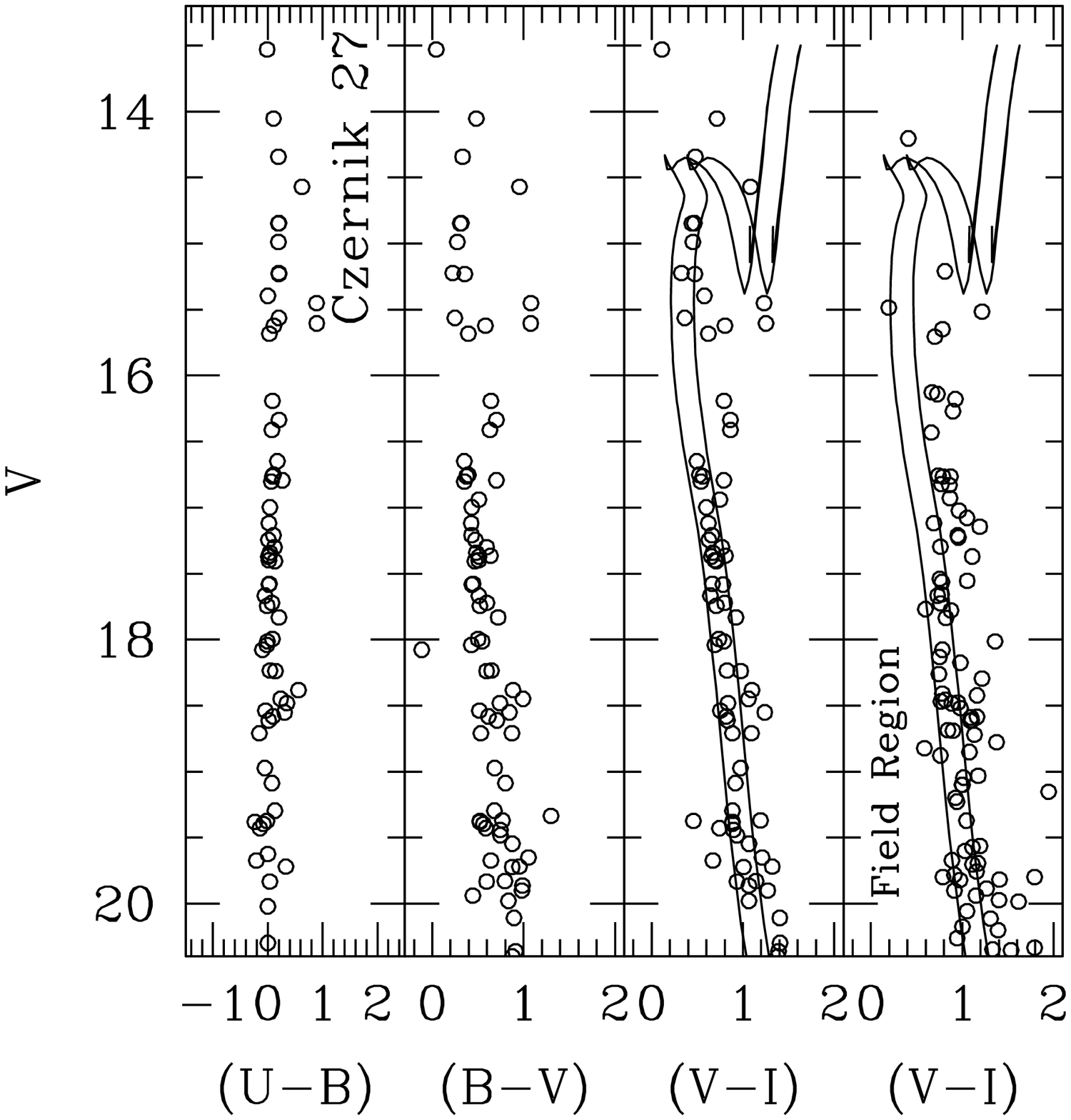}
}
\caption{The $V/(U-B)$, $V/(B-V)$ and $V/(V-I)$ CMDs for the clusters Be 24 and Cz 27 
using stars within the cluster radius. Stars outside the cluster radius are also plotted as field 
region stars in $V/(V-I)$ CMD. Solid lines show the blue and red envelope of the MS.}
  \label{cmd}
  \end{figure*}
$[Fe/H]=-14.316(\pm 1.919)\delta_{0.6}^2-3.557(\pm 0.285)\delta_{0.6}\\
+ 0.105(\pm 0.039)$.\\ 

The metallicity corresponding to the peak value for the $\delta_{0.6}$ distribution was calculated  
as $[Fe/H]= -0.025\pm 0.01$ dex for Be 24 and $[Fe/H]= -0.042\pm 0.01$ dex for Cz 27. In order to transform 
the $[Fe/H]$ metallicity  obtained from the photometry to the   mass   fraction   $Z$,   the  following relation 
(Mowlavi et al. (2012)) was used:\\

~~~~~~~~~~~~~~~~~~~$Z=\frac{0.013}{0.04+10^{-[Fe/H]}}$.\\

Theoretical isochrones corresponding to the metallicity $Z$, are selected for further estimation of the
astrophysical parameters for the clusters. We have found $Z$=0.01 to be suitable for the clusters.

    \begin{figure*}
    \centering
   \hbox{
   \includegraphics[width=9cm, height=10cm]{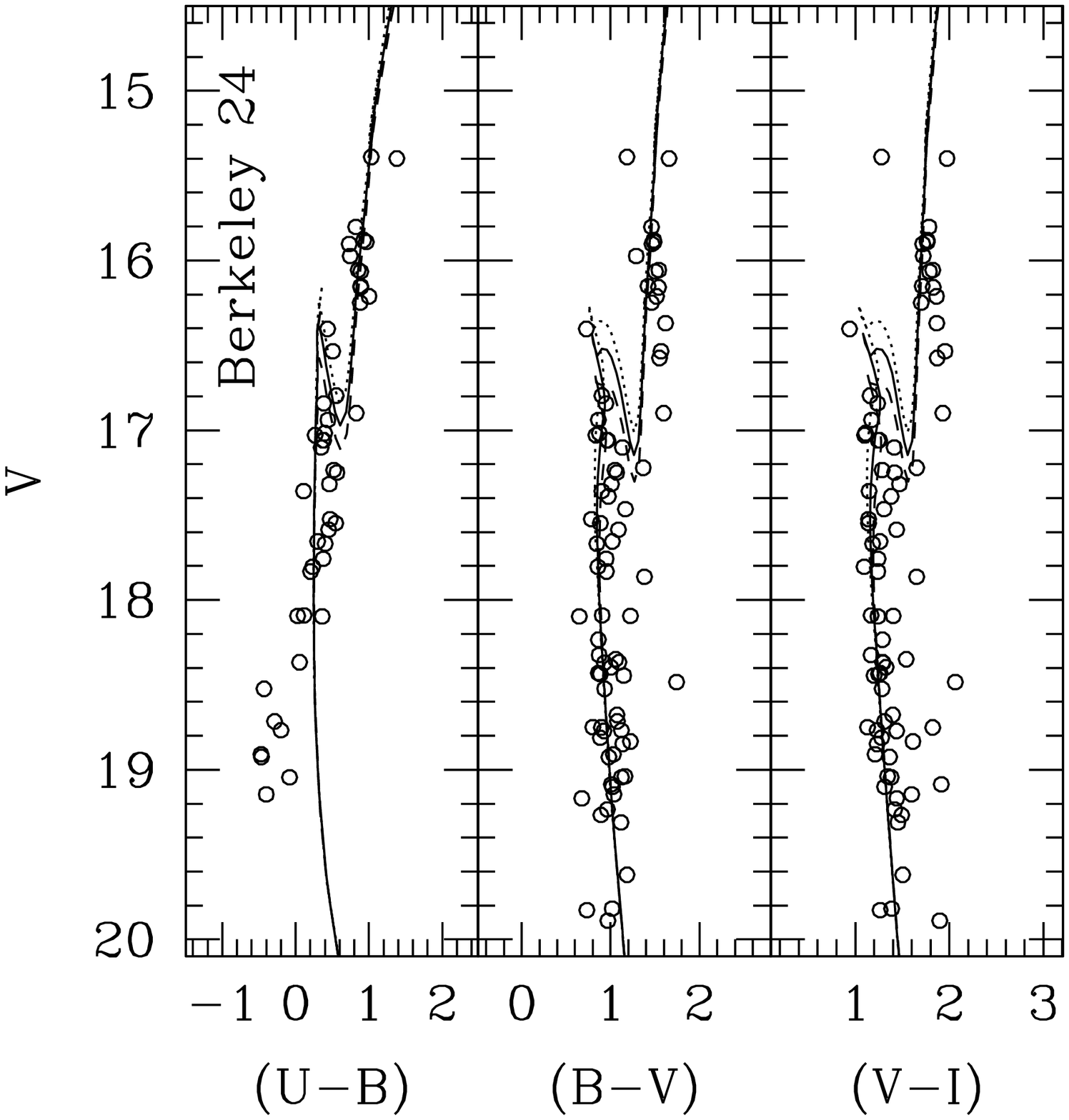}
   \includegraphics[width=9cm, height=10cm]{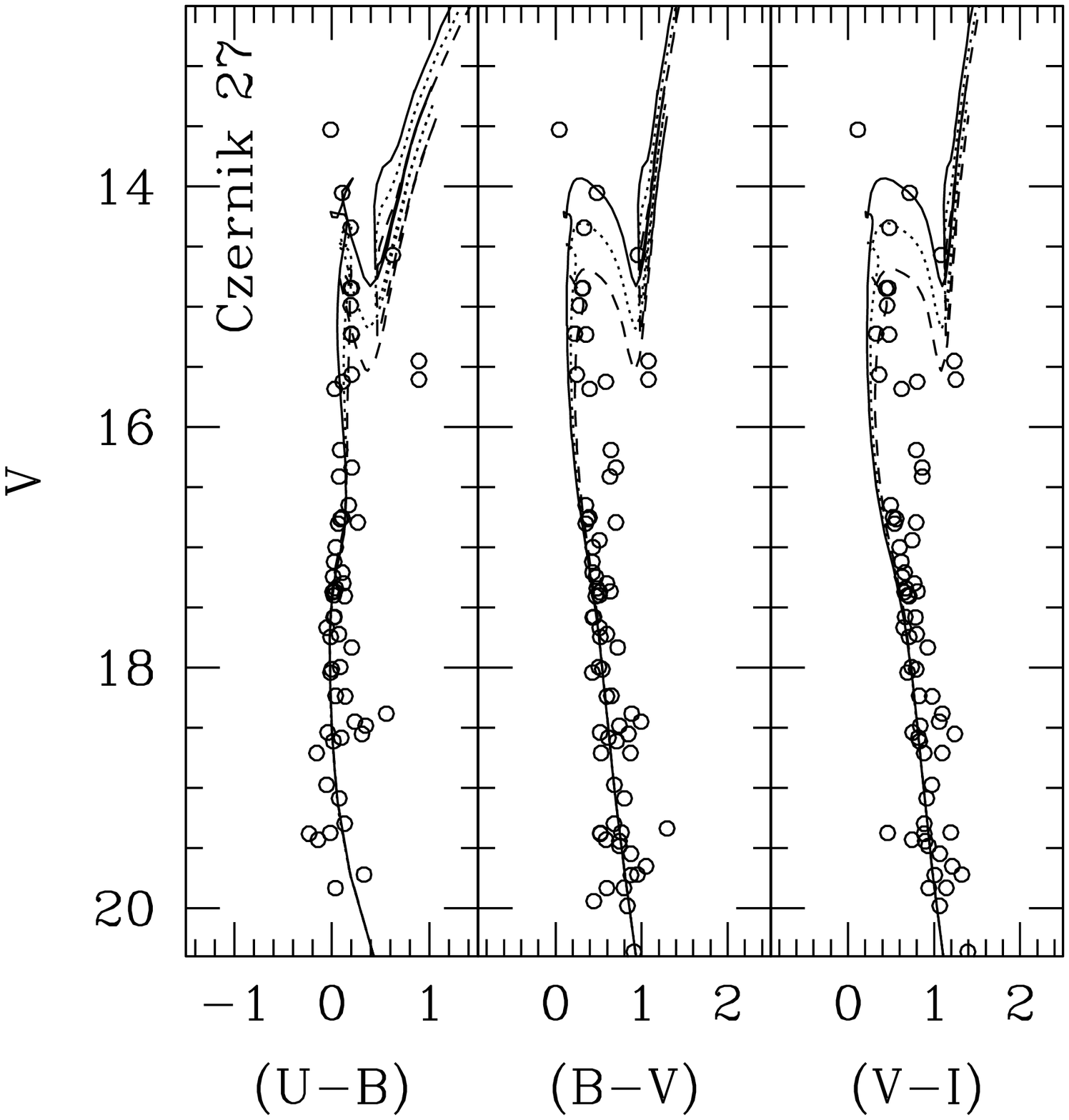}
}

\caption{The colour-magnitude diagram of the clusters under study. These stars are assumed
to be cluster members based on proper motion. The curves are the isochrones of (log(age) $=$  9.25 ,9.30
and 9.35) for the cluster Be 24 and (log(age) $=$  8.70 ,8.80 and 8.90) for the cluster Cz 27. These isochrones
are taken from Girardi et al. (2000).}
  \label{dist}
  \end{figure*}

\section{Ages and distances of the clusters}

The Colour-Magnitude diagram (CMD) is the most effective tool for the estimation of fundamental parameters such as age and distance 
for open star clusters. Identification of the main sequence in the cluster's CMD allows a model dependent mass, radius and distance
for each star to be determined. Since samples of stars in the cluster region on the sky are contaminated by the field star population,
it is necessary to distinguish the cluster sequence from field-star contamination for the proper analysis of the CMD. We identified
probable cluster members using VPD for both clusters. We have also constructed CMDs for stars in the field region as well. Stars 
present within circular area in VPD  for the clusters Be 24 and Cz 27 are considered as cluster region while stars outside this circular
area are considered field region stars. The $V, (U-B); V, (B-V)$ and $V, (V-I)$ CMDs of Be 24 and Cz 27 along with the $V, (V-I)$ CMDs 
of the corresponding field regions are shown in Fig.~\ref{cmd}. The CMD constructed using the stars within cluster radius for Be 24 shows 
main sequence extending down to $V\sim$20 mag except in the $V$, $(U-B)$ CMDs where it is only traced down to $V \sim$18.5 mag. For 
the cluster Cz 27, CMDs extends down to $V \sim$ 20 mag in $V/(B-V)$ and $V/(V-I)$. In $V, (V-I)$ CMDs, we have selected
the cluster members by defining the blue and red envelope around the main sequence, as shown in Fig \ref{cmd}. A star is 
considered as contaminant if it lies outside the strip in CMDs. Most of the field stars are separated out from the cluster
sequence using VPD and photometric criteria in both the clusters. Selected stars are used in LF/MF studies in the next section.

The fundamental parameters: reddening, metallicity, distance modulus and age of a cluster can be simultaneously determined by 
fitting the theoretical stellar isochrones to the observed colour-magnitude diagrams. In this case, we have used the reliable 
traditional methods, as discussed above, for the estimation of reddening and metallicity. In order to estimate distance and age 
of the clusters simultaneously, we have fitted theoretical isochrones given by Girardi et al. (2000) for the MS with $Z=0.01$. 
The  $V/(U-B), V/(B-V)$ and $V/(V-I)$ CMDs along with visually fitted isochrones are shown in Fig.~\ref{dist}. The detailed shape
and position of the different features in the CMD depend mostly on the age and metallicity of the clusters. There is larger
photometric scatter at the fainter end of the CMDs. This could be due to larger errors in the photometry or due to contamination 
of the CMD by field stars. The $E(B-V)$ values for both the clusters are taken from Sect.~3.4.2.

{\bf Be 24:} We superimpose isochrones of different age (log(age)$=$9.25, 9.30 and 9.35 with 
$Z = 0.01$ in $V/(U-B), V/(B-V)$ and $V/(V-I)$ CMDs. Using turn-off point, we have
found an age of $2\pm0.2$ Gyr. On average, we obtained a distance modulus of $(m-M)$ = $14.80\pm0.2$ mag.
The estimated distance modulus provides a heliocentric distance as $4.4\pm0.5$ kpc. The estimated 
distance for this cluster is higher than the 3.7 kpc derived by Koposov et al. (2008) using $2MASS$
$JHKs$ data, but similar to the value 4.4 kpc listed in Dias et al (2002). The Galactocentric coordinates 
are $X_{\odot} = -0.7$ kpc, $Y_{\odot} = 12.8$ kpc and 
$Z_{\odot} = -0.2$ kpc. The $Z$ coordinate indicates that this cluster is in the thin Galactic disk. The Galactocentric
distance of the cluster was calculated to be 12.8 kpc.

{\bf Cz 27:} As shown in Fig.~\ref{dist}, we have fitted the theoretical isochrones to $V/(U-B), V/(B-V)$ and $V/(V-I)$ 
CMDs. These isochrones of different age (log(age)=8.70, 8.80 and 8.90) and $Z$ = 0.01 have been superimposed on the CMDs.
We have found an age of $0.6\pm0.1$ Gyr. The distance modulus $(m-M)$ = 14.30 mag provides a heliocentric 
distance as $5.6\pm0.2$ kpc which is in good agreement with the value 5.8 kpc listed in Dias et al (2002). The Galactocentric 
distance is determined as 14.1 kpc, which is determined by assuming 8.5 kpc as the 
distance of the Sun to the Galactic center. The Galactocentric coordinates are estimated as $X_{\odot} = -0.7$ kpc,
$Y_{\odot} = 14.0$ kpc and $Z{\odot} = 0.5$ kpc.

The present age estimated for Be 24 is similar to the age derived by Ortolani et al. (2005) as 2.2 Gyr.
For Cz 27, Piatti et al. (2010) estimated an age of 0.7 Gyr, which agrees well within error to the present estimate.

\begin{figure}
    \centering
   \hbox{ 
   \includegraphics[width=4.5cm, height=4.5cm]{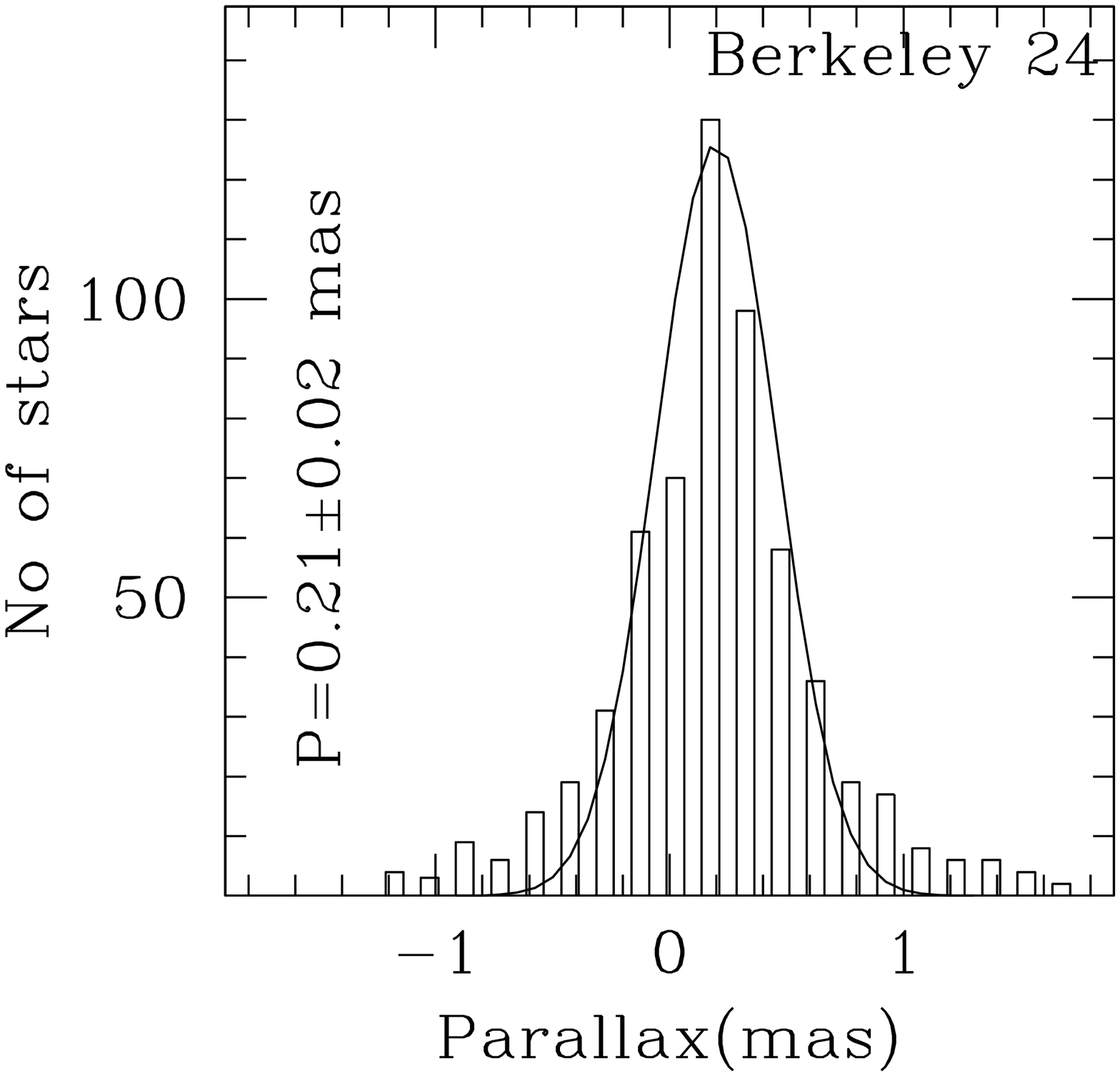}
    \includegraphics[width=4.5cm, height=4.5cm]{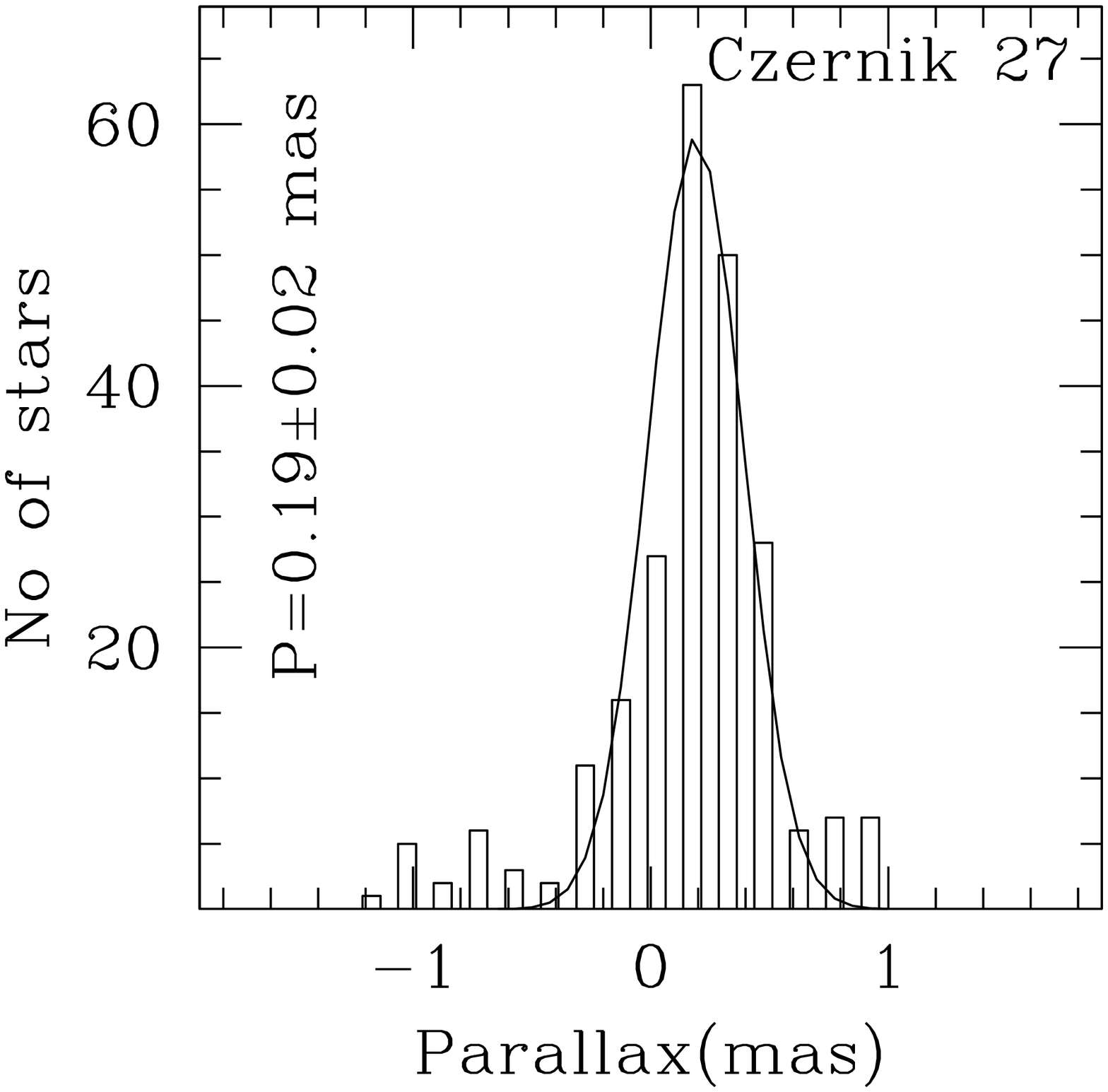}
}
\caption{Parallax histograms of 0.15 mas bins for our candidate clusters. The Gaussian function fit to 
the central bins provides mean parallax.}
  \label{dist_parallax}
  \end{figure}

We used parallax of cluster stars taken from GDR2 for both the clusters. We constructed histograms of 0.15 mas bins as 
shown in Fig \ref{dist_parallax} using probable members selected from VPD. Mean parallax is estimated as $0.21\pm0.02$ mas and $0.19\pm0.02$ mas
for clusters Be 24 and Cz 27 respectively and corresponding distances are $4.7\pm0.5$ kpc and $5.3\pm0.5$ kpc. Estimated 
values of distance from observed data and from GDR2 are very good in agreement with each other for both clusters.

\subsection{Optical and near-IR CMDs}
 
\begin{figure*}
    \centering
   \hbox{
   \includegraphics[width=9cm, height=8cm]{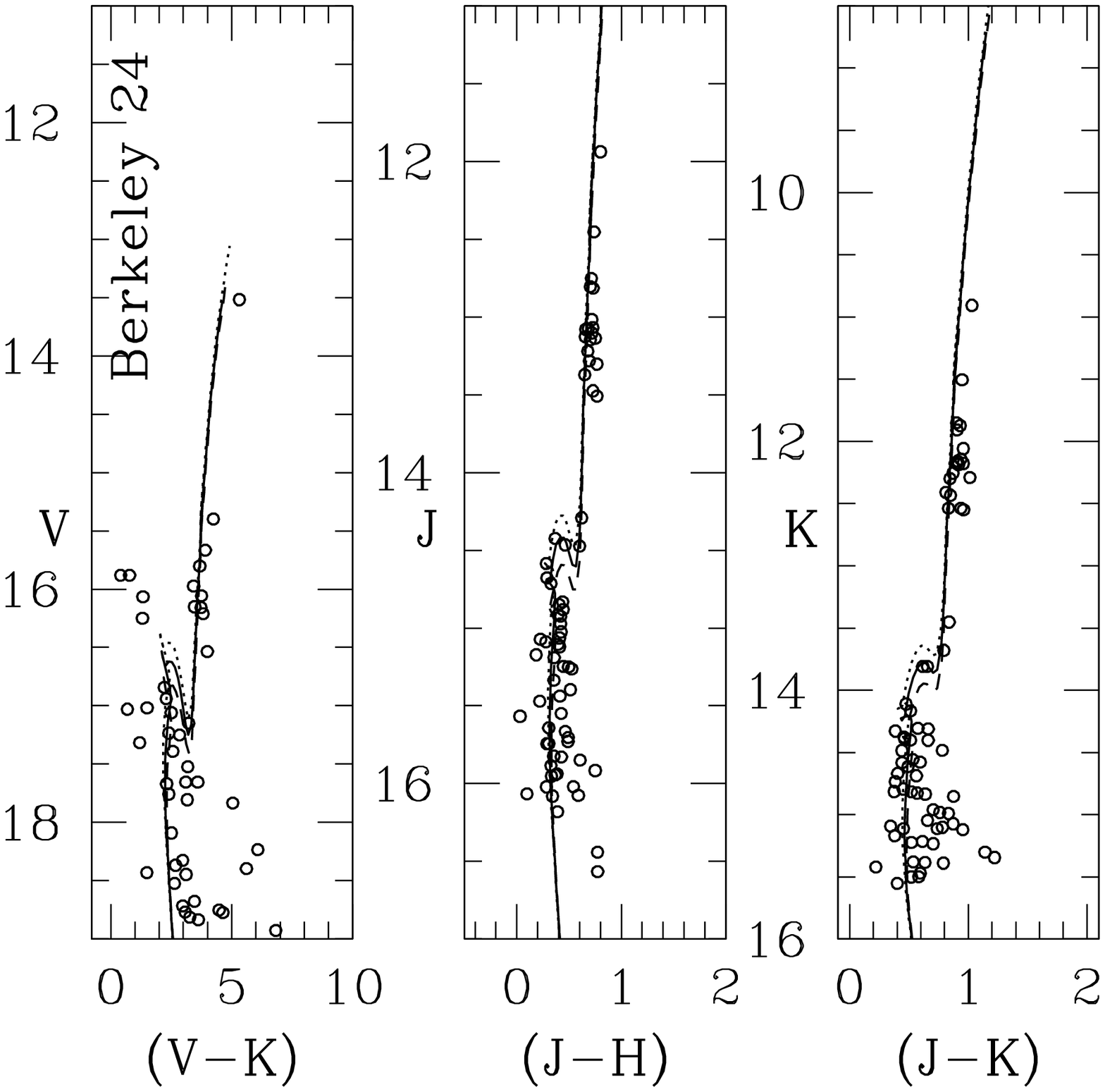}
   \includegraphics[width=9cm, height=8cm]{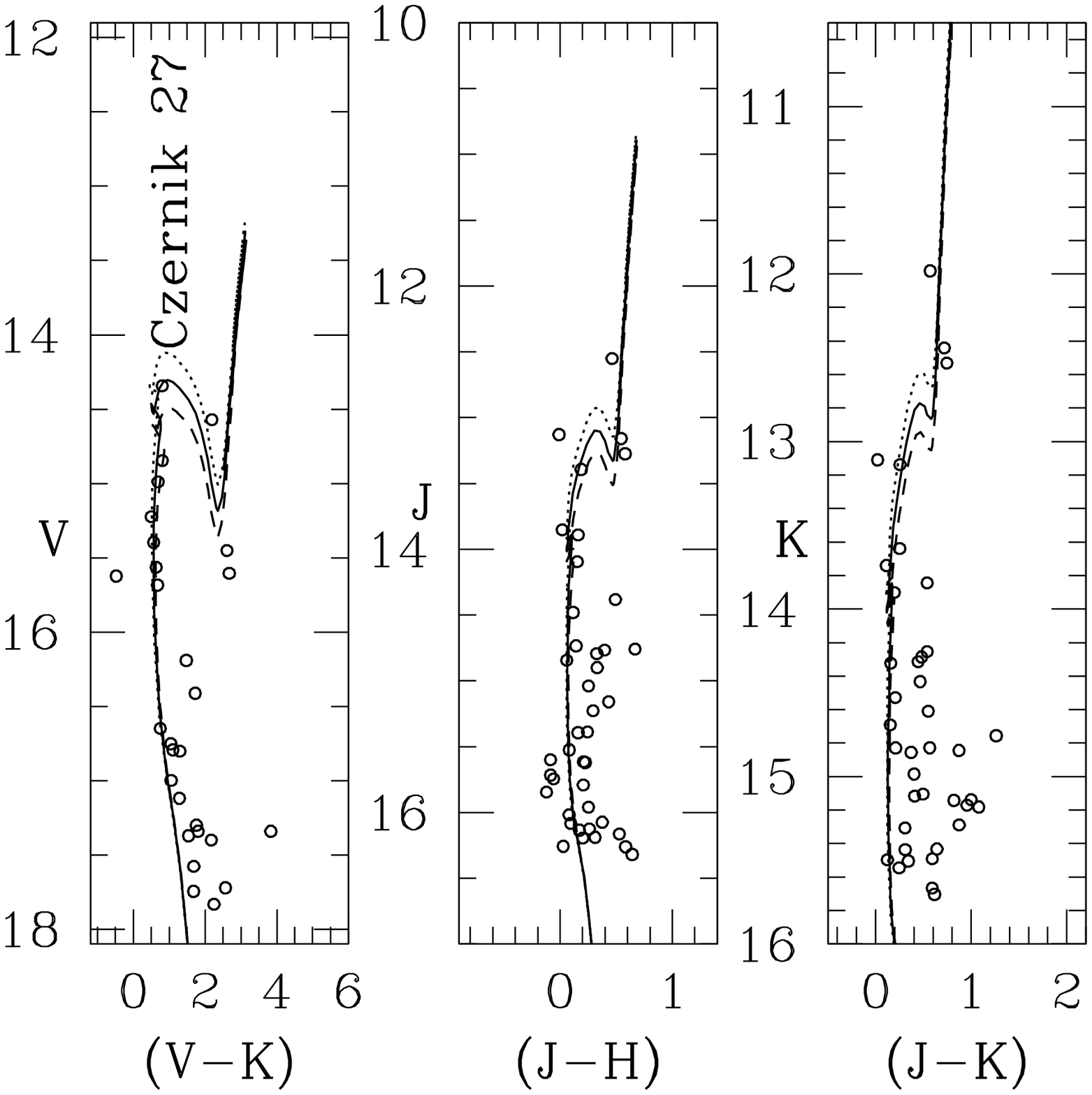}
}
\caption{Same as Fig.~\ref{dist} of optical and near-IR colour-magnitude diagrams for Be 24 and Cz 27.}
  \label{cmd_jk}
  \end{figure*}

Using both, optical and near-IR data we have re-determined distance and age of the clusters. We have 
plotted $V$ versus $(V-K)$ and $K$ versus $(J-K)$ CMDs, which is shown in Fig.~\ref{cmd_jk}. The theoretical 
isochrones given by Girardi et al. (2000) for $Z =$ 0.01 of log(age)=9.30 and 8.80 have been overplotted in 
the CMDs of Be 24 and Cz 27 respectively. The apparent distance moduli $(m - M)_{V, (V-K)}$ and 
$(m-M)_{K, (J-K)}$ turn out to be 14.0$\pm$0.3 and 12.0$\pm$0.3 mag for the cluster Be 24 and 14.0$\pm$0.3
and 12.1$\pm$0.3 mag for the cluster Cz 27. Using the reddening values estimated in \S\ref{sec:extir}, we 
have derived a distance of 4.3$\pm$0.4 kpc for Be 24 and 5.4$\pm$0.3 kpc for Cz 27. Both age and distance determination 
for the clusters are thus in agreement with the estimates using optical data. However, the scatter is larger due to the 
large errors in the $JHK$ magnitudes.

\section{Luminosity and mass function study} \label{sec:mfu}

To study the luminosity function (LF) and mass function (MF), the first necessary step is to remove 
the field star contamination from the sample of stars in the cluster region. A statistical field star subtraction method has 
been used by assuming that the field stars within the cluster and surrounding areas are distributed 
in a similar way (Wilner \& Lada 1991; Phelps \& Janes 1993 and Sagar \& Griffiths 1998). A brief 
description of the procedure applied here is described in Section 5. 

\subsection{Completeness of the CCD photometry}

\begin{figure}
     \centering
     \includegraphics[width=6cm, height=6cm]{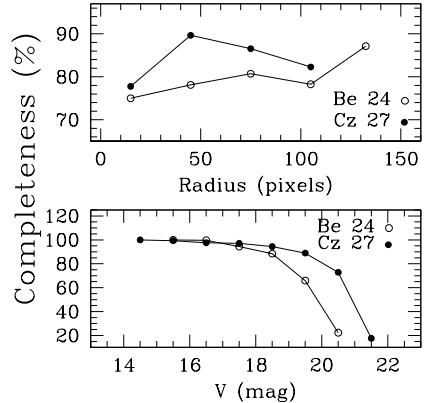}
 \caption{Variation of completeness factor versus $V$ magnitude and radius for Be 24 and Cz 27.}
   \label{cf_var}
   \end{figure}

Photometric data may be incomplete due to the stellar crowding as well as inefficiency of data reduction programmes. The 
incompleteness correction is very necessary to compute luminosity function of the stars in the cluster. Completeness 
corrections were determined by running artificial star experiments on the data. The {\bf ADDSTAR} routine in {\bf DAOPHOT II} 
was used to determine Completeness factor (CF). We created several images by adding artificial stars to the original $V$ image.
Stars were added at same geometrical positions in $I$ band image. Stars found in both $V$ and $I$ band were considered as
real detection. In order to avoid appreciable increase in the crowding, we have 
randomly added only 10 to $15\%$ of actually detected stars of known magnitude and position into the original images. The 
luminosity distribution of artificial stars is chosen in such a way that large number of stars are inserted into the fainter 
magnitude bins. This is the range of magnitudes where significant losses due to incompleteness are expected 
(Bellazzini et al. 2002).
Detailed information about this experiment is given by Yadav \& Sagar (2002) and Sagar \& Griffiths (1998). 
The added star images are re-reduced using the same method, which was 
adopted for the original images. The ratio of recovered to added stars in different magnitude bins gives the CF. We have
estimated CF in core, halo and overall region for both the clusters. The CF derived in this way are listed in  Table~\ref{tab:CF} for
the clusters Be 24 and Cz 27. As expected the incompleteness of the data increases with increasing magnitude and increasing stellar
crowding. Table~\ref{tab:CF} shows the value of CF in different magnitude bins and in different regions. The variation of CF
versus $V$ magnitude and radius for both the clusters Be 24 and Cz 27 are shown in Fig \ref{cf_var}. The value 
of CF is $\sim$ 90\% and $\sim$ 65\% at 20 mag in $V$ band for clusters Be 24 and Cz 27 respectively. CF versus radius curve 
indicates that the CF is more than 75\% for both the clusters within cluster extent. 

\subsection{Luminosity function}

\begin{figure}
    \centering
   \hbox{ 
   \includegraphics[width=4cm, height=4cm]{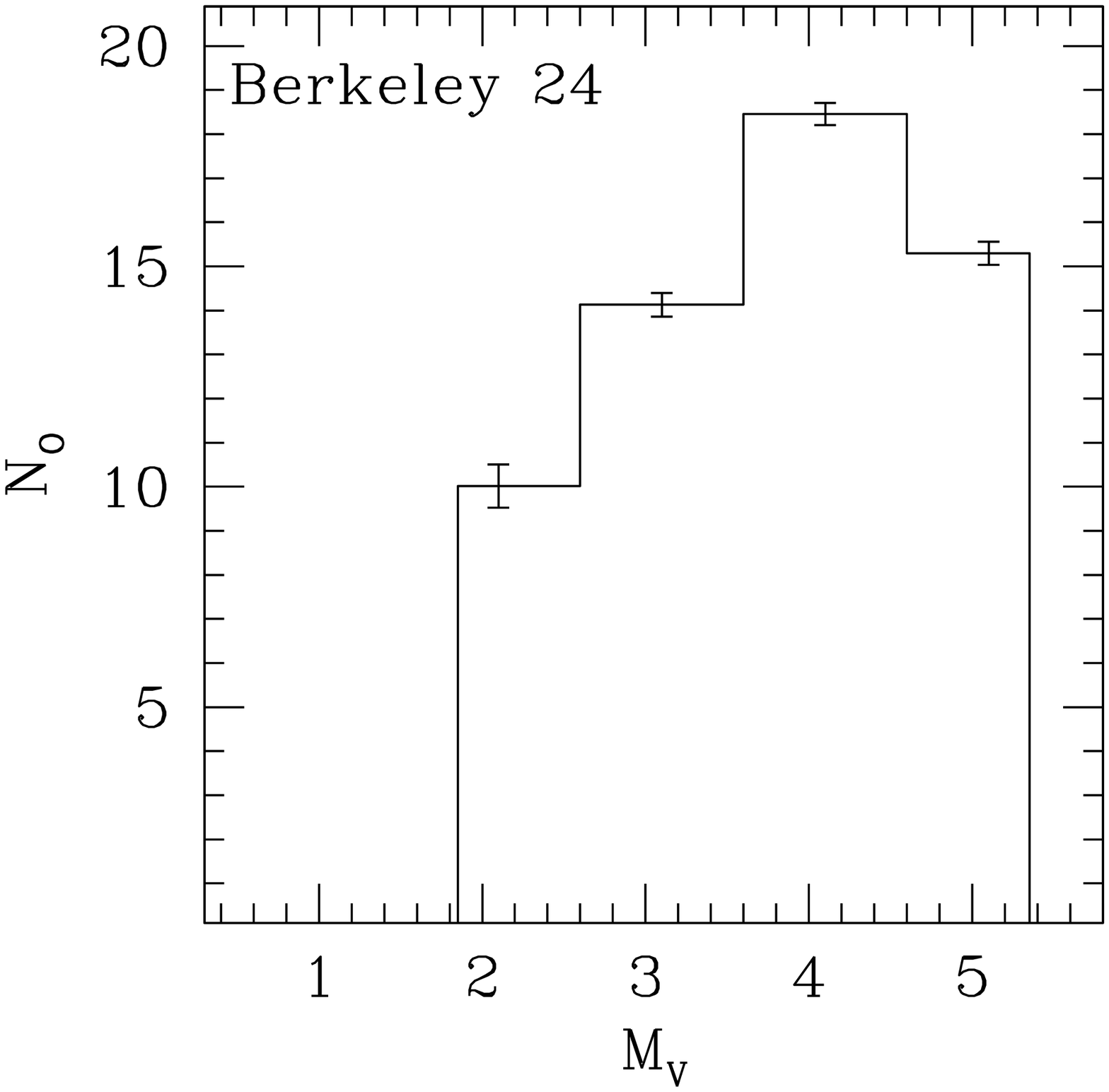}
    \includegraphics[width=4cm, height=4cm]{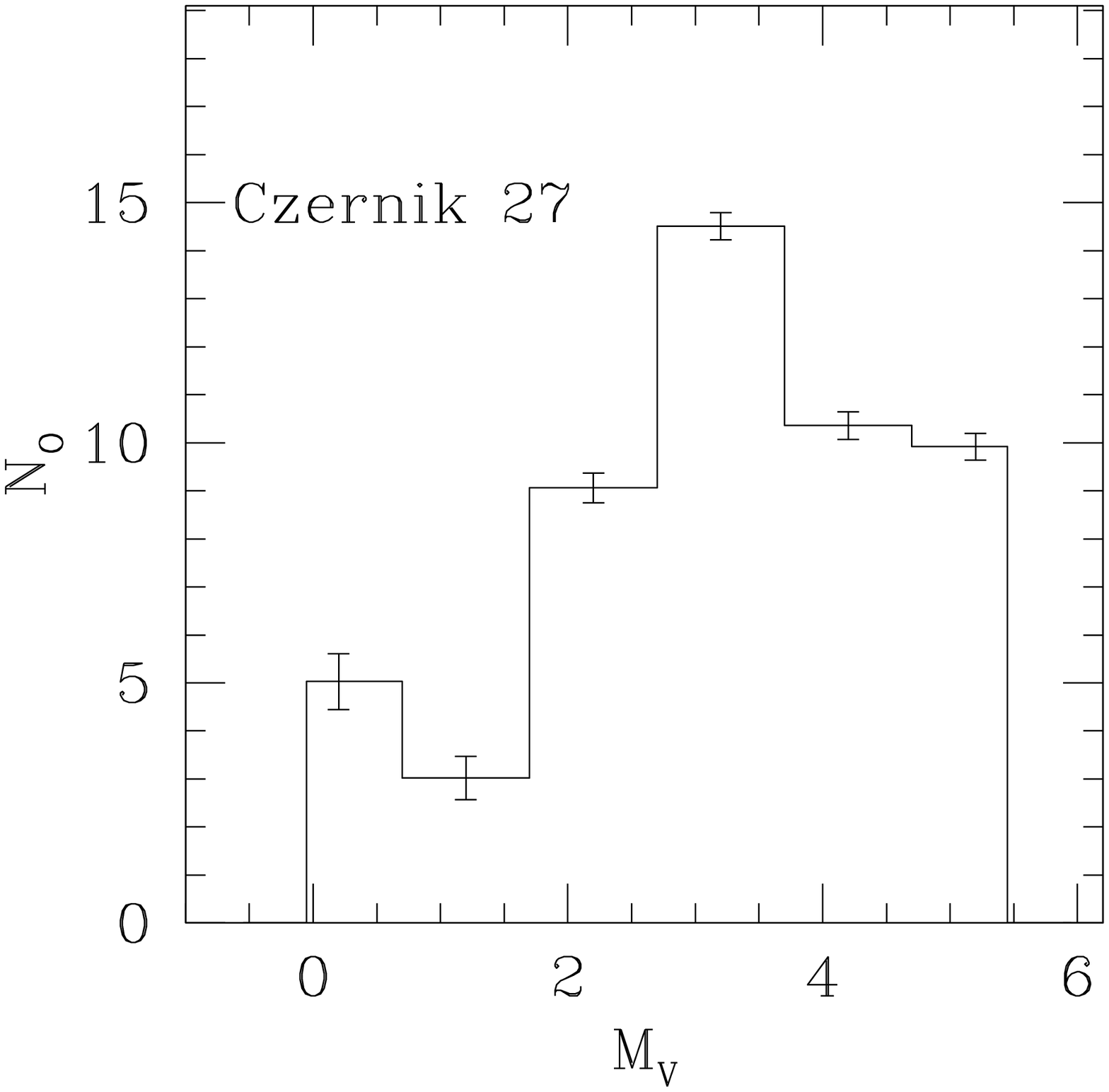}
}
\caption{Luminosity function of clusters Be 24 and Cz 27 under present study.}
  \label{lf}
  \end{figure}

We have used the $V/(V-I)$ colour magnitude diagram to construct the luminosity function (LF) for target clusters. In $V/(V-I)$ 
diagram, we used probable cluster members selected from VPD as shown in Fig \ref{pm_cmd}. It is necessary to clearly separate field 
stars from the cluster sequence for the construction of correct LF. So, we have collectively used proper motion and photometric criteria
for the selection of  probable cluster members. The photometric criterion adopted in Sec. 5 by plotting the blue and red envelope 
along the MS was used for selection of cluster members in the $V$ versus $(V-I)$ CMD of cluster field region (see Fig.~\ref{cmd}). 
Stars are counted within this envelope, for both cluster and
field region. The difference between the counts in two fields after accounting for the difference in area between the cluster and 
field regions will be the observed cluster LF. For the construction of LFs, first we have transformed the apparent $V$ magnitudes 
into the absolute magnitudes by using the distance modulus of the clusters. Then we have built the histogram of LFs for the clusters 
as shown in Fig.~\ref{lf}. The interval of 1.0 mag was chosen so that there would be enough stars per bin for good statistics. The 
histogram shows that a dip is found at $M_{V}$=1.0 mag for the cluster Cz 27. The LF for the cluster Be 24 rises steadily up 
to $M_{V}$=4.0 mag. 

\subsection{Mass function}

The LF can be transformed into the mass function (MF) using a mass-luminosity relation.
Since we could not obtain an empirical transformation, so we rely on theoretical models. 

\begin{figure}
    \centering
   \hbox{ 
    \includegraphics[width=4cm, height=4cm]{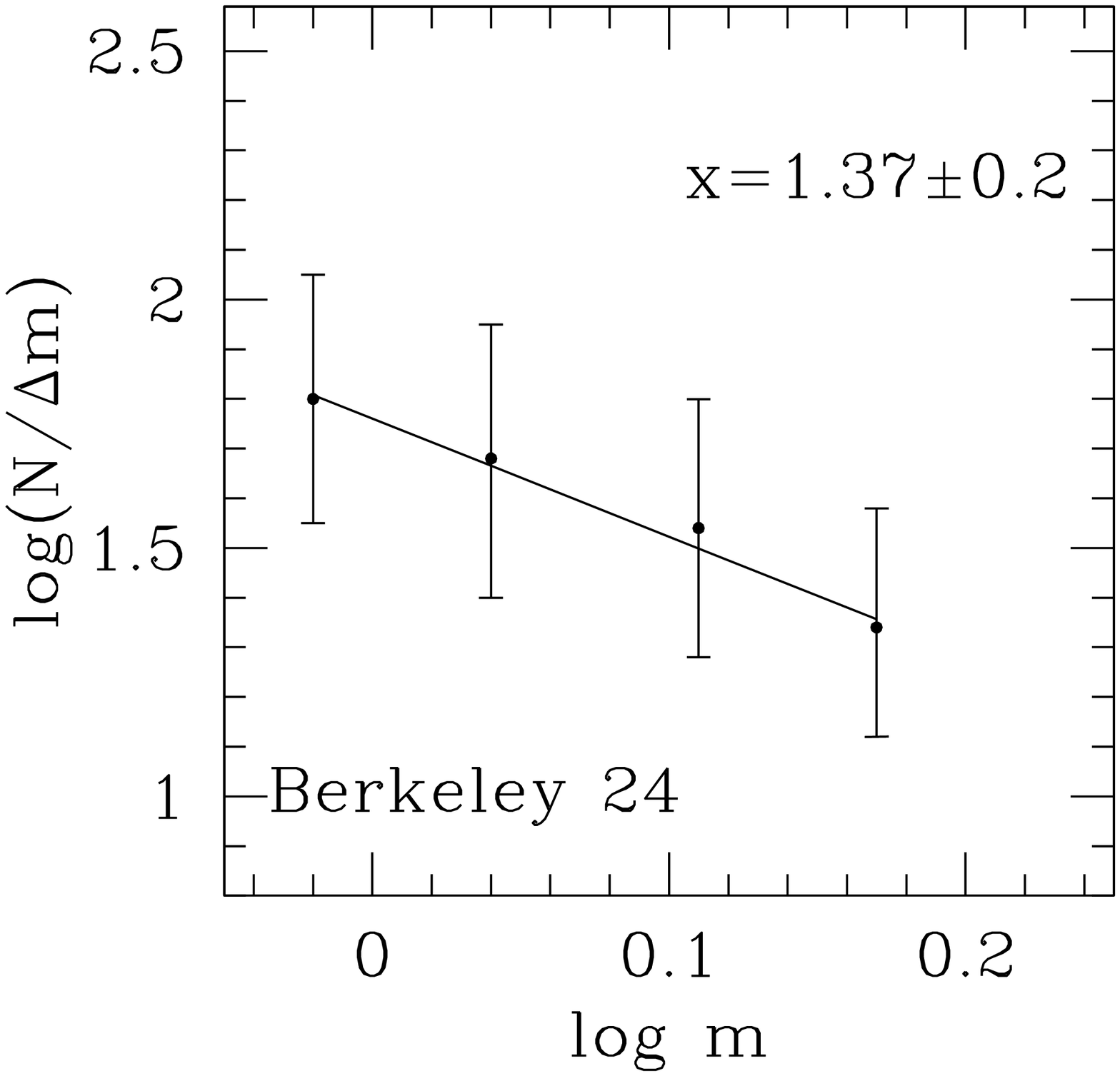}
   \includegraphics[width=4cm, height=4cm]{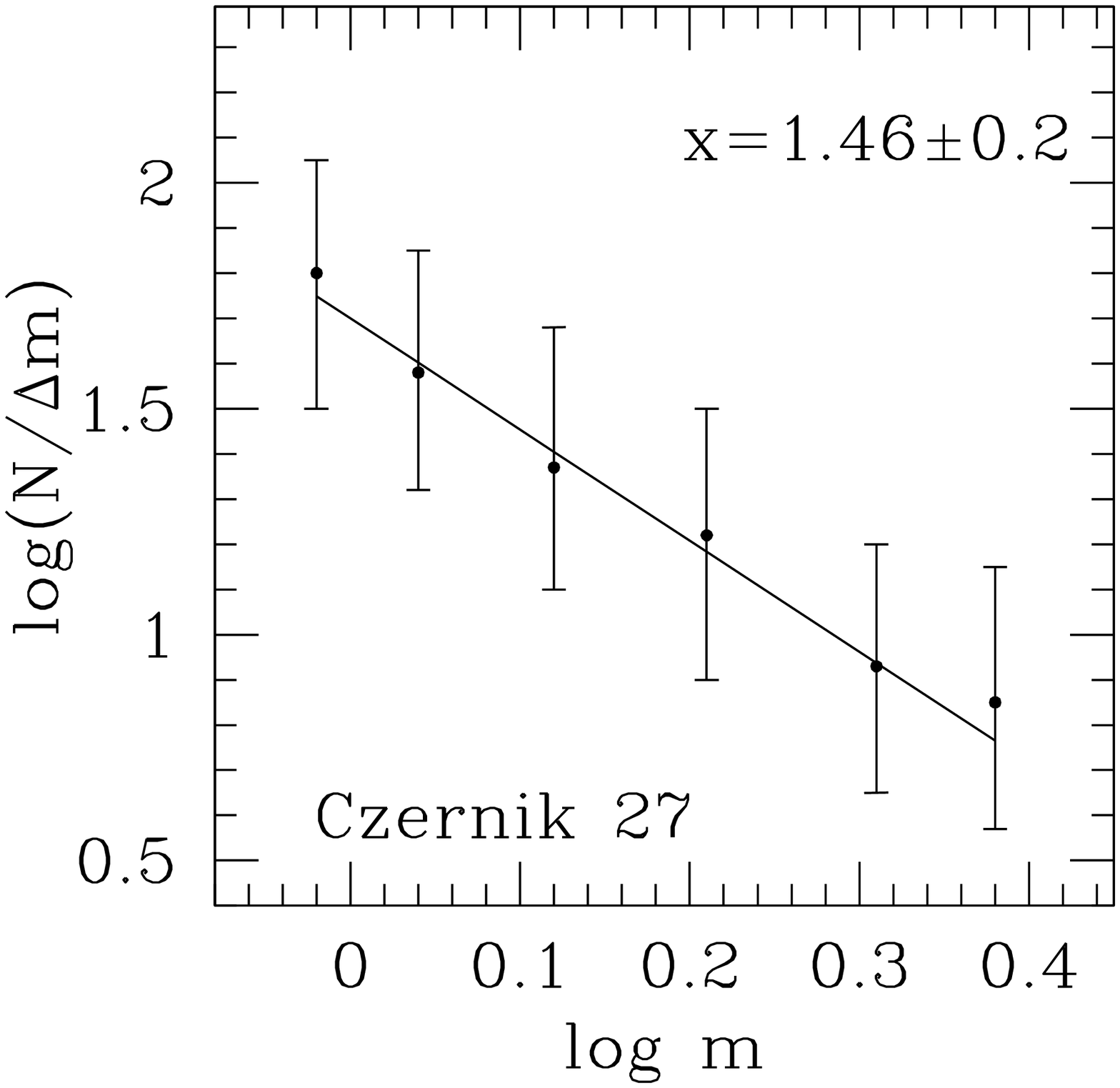}
}
\caption{Mass function for Be 24 and Cz 27 derived using Girardi et al. (2000) isochrones. The error bars
represent $\frac{1}{\sqrt(N)}$.}
  \label{mass}
  \end{figure}

Using the cluster parameters derived in this paper and theoretical models given by Girardi et al. (2000), we have converted
the LF into an MF. To convert LFs derived in Sec 6.2 into MFs, we divide the cluster members by the mass interval
of the magnitude bin under consideration. The value of mass interval was obtained from the mass-luminosity relation 
derived from the appropriate isochrone. The resulting MF for both the clusters are shown in Fig.~\ref{mass}. The mass 
function slope can be derived from the linear relation\\

${\rm log}\frac{dN}{dM} = -(1+x) \times {\rm log}(M)+${}constant\\

using the least-squares solution. In the above relation, $dN$ represents the number of stars in a mass bin $dM$ 
with central mass $M$ and $x$ is the mass function slope. The Salpeter (1955) value for the mass function slope 
is $x=1.35$. 

The derived mass function slope is x=$1.37\pm0.2$ and $1.46\pm0.2$ for the clusters Be 24 and Cz 27 respectively. The 
overall MF slope for both the clusters are in good agreement with the Salpeter value (x=1.35) within errors. The mass is
calculated by considering overall mass function slope with in the mass range 0.9~-~1.6 $M_{\odot}$ for Be 24 and
0.9~-~2.4 $M_{\odot}$ for Cz 27. Total mass was estimated as $\sim$80 $M_{\odot}$ and $\sim$84 $M_{\odot}$ for 
clusters Be 24 and Cz 27 respectively.

\begin{table}
\centering
\caption{Variation of completeness factor (CF) in the $V$, $(V-I)$ 
      diagram with the MS brightness.}
\begin{tabular}{c|cccccc}
\hline
$V$ &\multicolumn{3}{c}{Be 24} &\multicolumn{3}{c}{Cz 27}\\
&whole&core&halo&whole&core&halo\\
mag &&&&&&\\
\hline
14-15&99.99& 99.99& 99.99&99.99& 99.99& 99.99\\
15-16&99.35& 90.90& 97.92&99.99& 99.99& 99.99\\
16-17&97.56& 81.81& 97.91&99.65& 94.11& 95.18\\
17-18&97.15& 93.54& 90.56&94.39& 88.00& 95.04\\
18-19&94.42& 93.11& 93.49&88.48& 86.95& 90.68\\
19-20&88.97& 92.25& 92.44&65.94& 49.05& 72.26\\
\hline
\end{tabular}
\label{tab:CF}
\end{table}


There are many MF studies available in the literature using open star clusters in the Milky Way.
Piskunov et al. (2004) studied 5 young open star clusters and found that stellar mass function of these clusters
are well represented with a power law, very similar to Salpeter value within the uncertainties. MF study of 
Phelps et al. (1993) using a sample of seven star clusters also conclude the Salpeter type MF slope in 
these clusters. Sanner et al. (2001), Sagar et al. (2001), Yadav et al. (2002, 2004) also found consistent with 
the Salpeter value.

\begin{table*}
\centering
\caption{Derived fundamental parameters of the clusters under study. $R_{GC}$ is the
Galactocentric distance while $X_{\odot}$, $Y_{\odot}$ and $Z_{\odot}$ are the Galactocentric 
coordinates of the clusters. The coordinate system is such that the Y-axis connects the
Sun to the Galactic Center, while the X-axis is perpendicular to that. $Y_{\odot}$ is
positive towards the Galactic anticentre, and $X_{\odot}$ is positive in the first and
second Galactic quadrants (Lynga 1982).}

\begin{center}
\small
\begin{tabular}{cccccccccc}
\hline
Name& Radius  & $E(B-V)$ &Distance & $X_{\odot}$ & $Y_{\odot}$ & $Z_{\odot}$ & $R
_{GC}$  & Age   \\
  &(arcmin)   &(mag)     &(kpc)    &(kpc)    &(kpc)   &(kpc)   &(kpc)  &(Gyr)\\
\hline
Be 24 & 2.7 & $0.45\pm0.05$ & $4.4\pm0.5$ & -0.7 & 12.8 & -0.2 & 12.8 & $2\pm0.2$  \\
Cz 27 & 2.3 & $0.15\pm0.05$ & $5.6\pm0.2$ & -0.7 &  14.0 & 0.5 & 14.1 & $0.6\pm0.1$  \\
\hline\hline
\end{tabular}
\label{final_para}
\end{center}
\end{table*}


\section{Dynamical state of the clusters} \label{sec:mse}

\begin{figure}
    \centering
   \hbox{ 
    \includegraphics[width=4cm, height=4cm]{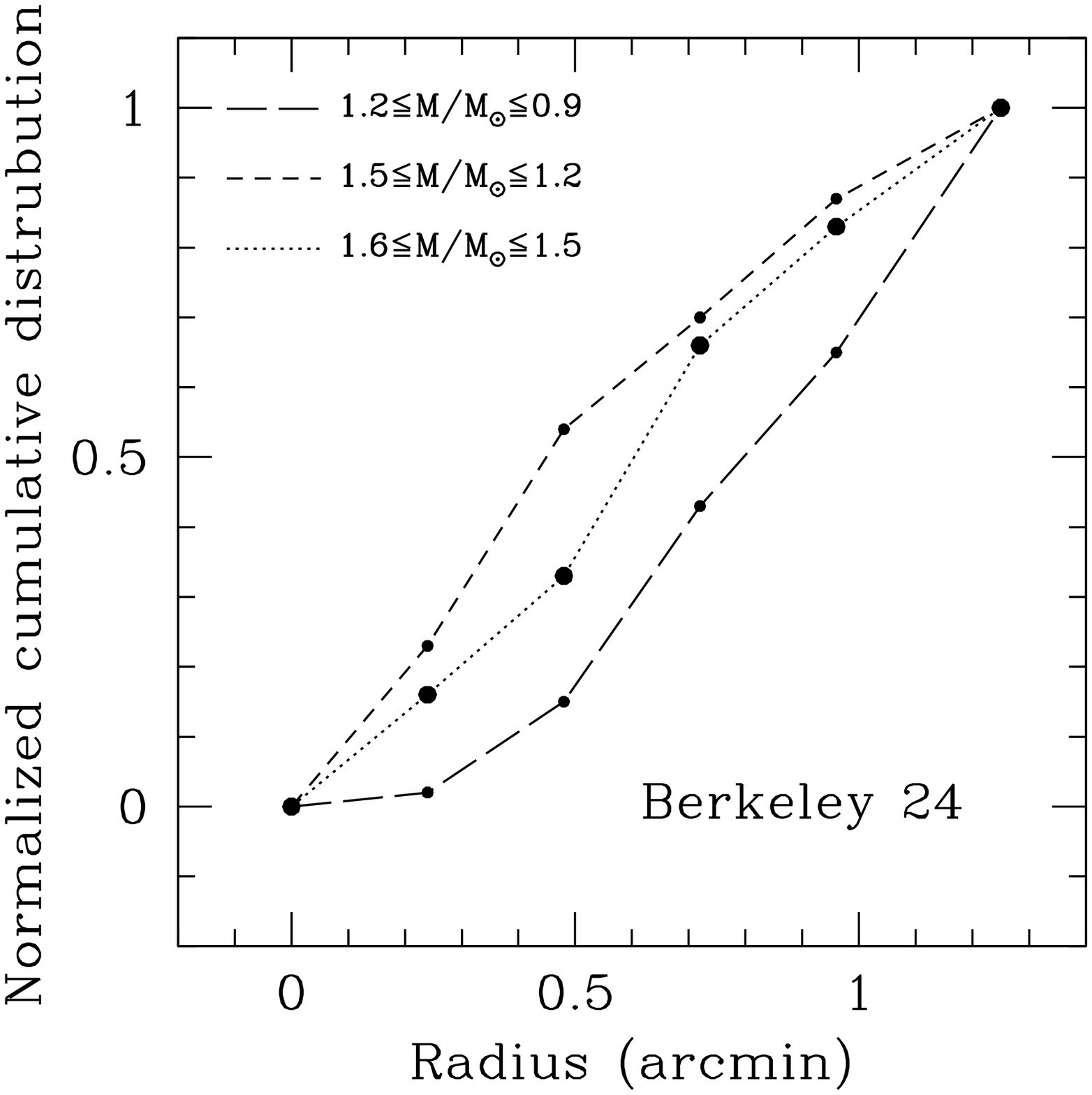}
   \includegraphics[width=4cm, height=4cm]{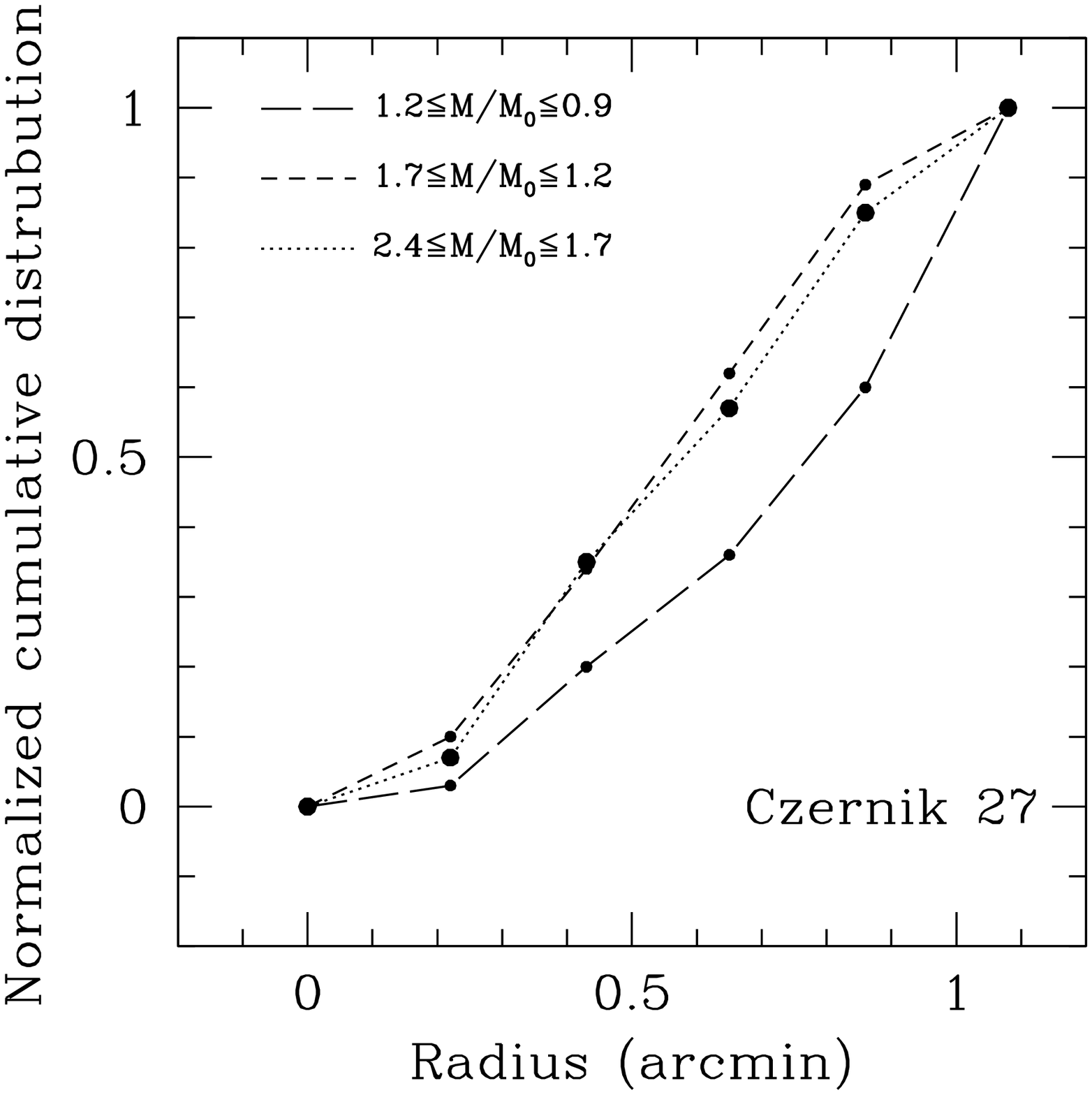}
}
\caption{The cumulative radial distribution of stars in various mass range.}
  \label{mass_seg}
  \end{figure}

To study the effect of mass segregation on the clusters, we plot the cumulative radial stellar distribution of stars for 
different masses in Fig.~\ref{mass_seg}. To bring out the mass segregation effect, we have divided the 
main sequence stars in three mass ranges, 1.6$\le\frac{M}{M_{\odot}}\le$~1.5, 1.5$\le\frac{M}{M_{\odot}}\le$~1.2 and
1.2$\le\frac{M}{M_{\odot}}\le$~0.9 for Be 24 and 2.4$\le\frac{M}{M_{\odot}}\le$~1.7,
1.7$\le\frac{M}{M_{\odot}}\le$~1.2 and 1.2$\le\frac{M}{M_{\odot}}\le$~0.90 for Cz 27. To study the cluster dynamical
evolution and mass-segregation effect, we selected probable members based on VPD and CMD of the clusters Be 24 and Cz 27.
The cumulative distributions shown in Fig.~\ref{mass_seg} are also corrected for the incompleteness as listed in Table~\ref{tab:CF} 
and field star contamination. This figure clearly exhibits mass-segregation effect, in the sense that massive stars gradually sink 
towards the cluster center than the fainter stars. To check whether these mass distributions represent the same kind of distribution
or not, we have performed Kolmogorov-Smirnov $(K-S)$ test. This test indicates that the confidence level of mass-segregation effect 
is 92 $\%$ for Be 24 and 90 $\%$ for  Cz 27.

Furthermore, it is important to know weather the effect mass-segregation is due to dynamical evolution or imprint of
star formation or both. In the lifetime of star clusters, encounters between its member stars gradually lead to an 
increased degree of energy equipartition throughout the clusters. In this process the higher mass cluster members 
accumulate towards the cluster center and transfer their kinetic energy to the more numerous lower-mass stellar component,
thus leading to mass segregation. 

\subsection{The relaxation time of clusters}

The time scale in which a cluster will loose all traces of its initial conditions is well represented by 
its relaxation time $T_{E}$. The relaxation time is the characteristic time-scale for a cluster to reach some 
level of energy equipartition. The relaxation time given by Spitzer \& Hart (1971) stated that;\\

~~~~~~~~~~$T_{E} = \frac{8.9 \times 10^{5} N^{1/2} R_{h}^{3/2}}{<m>^{1/2}log(0.4N)}$\\

\noindent
where $N$ is the number of cluster members, $R_{h}$ is half-mass radius of the cluster 
and $<m>$ is mean mass of the cluster stars (Spitzer 1971). We identified 59 and 54 stars
as probable cluster members for Be 24 and Cz 27 respectively, based on proper motion and
photometric criteria. The value of $<m>$ is found as 1.35 $M_{\odot}$ and 1.55 $M_{\odot}$ for clusters
Be 24 and Cz 27 respectively. The value of $R_{h}$ has been assumed as half of the cluster radius 
derived by us. Using the above relation, we have estimated the dynamical relaxation time $T_E$ as 9.5 
and 8.0 Myr for Be 24 and Cz 27 respectively. A comparison of cluster age with its relaxation time 
indicates that relaxation time is smaller than the age of these clusters. Therefore, we conclude that 
both the clusters are dynamically relaxed.

%
%
%
%
\section{Conclusions} \label{sec:con}

We have studied the two open star clusters Be 24 and Cz 27 using $UBVI$ CCD, $2MASS$ $JHK$ and $GAIA~~DR2$ data. The results
are summarized in Table~\ref{final_para}. The main findings of our analysis are as follows:

\begin{enumerate}

\item The radii of the clusters are obtained as 2\farcm7 and 2\farcm3 which corresponds to 3.4 and 
3.2 pc, respectively, at the distance of the clusters Be 24 and Cz 27.\\

\item From the two colour diagram, we have estimated $E(B-V) = 0.45\pm0.05$ mag for Be 24 and $0.15\pm$0.05 mag
for Cz 27. The $JHK$ data in combination with the optical data provide $E(J-K) = 0.23\pm$0.03 mag and 0.06$\pm$0.02 mag
while $E(V-K) = 1.23\pm$0.02 mag and $E(V-K) = 0.33\pm$0.01 mag for Be 24 and Cz 27, respectively. Hence, our analysis 
indicates that interstellar extinction law is normal towards both clusters.\\  

\item The metallicities of the clusters obtained from $UBV$ photometric data are found to be [Fe/H]= $-0.025\pm0.01$ dex 
and $-0.042\pm0.01$ dex for Be 24 and Cz 27, respectively. These are the first metallicity measurements for these two clusters.\\

\item Distances to the clusters Be 24 and Cz 27 are determined to be 4.4$\pm$0.5 and 5.6$\pm$0.2 kpc, respectively. 
These distances are supported by the distance values derived using parallax and optical and near-IR data. 
Ages of 2.0$\pm$0.2 Gyr and $0.6\pm0.1$ Gyr are estimated for Be 24 and Cz 27 by comparing with the isochrones for $Z =$ 0.01
given by Girardi et al. (2000).\\

\item The mean proper motion was estimated to be $1.25\pm0.09$ mas yr$^{-1}$ and $1.40\pm0.05$ mas yr$^{-1}$ 
      for the clusters Be 24 and Cz 27, respectively. \\

\item The luminosity function is determined by considering probable cluster members based on VPD and photometric criteria.
For the cluster Cz 27 a dip is found at $M_{V}$=1.0 mag and then it rises. The reason for the presence of dip a in the main sequence of 
the cluster is not known.\\

\item The overall mass function slopes $x = 1.37\pm0.2$ and $1.46\pm0.2$ are derived for Be 24 and Cz 27 by 
considering the corrections of field star contamination and data incompleteness.\\

\item Evidence for the mass-segregation effect was found for both clusters using probable cluster members. The K-S test 
shows that the confidence level of mass-segregation effect is 92 $\%$ and 90 $\%$ for Be 24 and Cz 27 respectively. Dynamical relaxation
time indicates that both the clusters are dynamically relaxed. This may be due to the dynamical evolution.\\

\end{enumerate}

{\bf ACKNOWLEDGMENT}\\

The authors thank the anonymous referee for useful comments that improved the scientific content
of the article significantly. We also acknowledge ARIES for great support during observations. Work at PRL is
supported by the Dept. of Space. The Cosmic Dawn Center is funded by the DNRF. This publication has made 
use of data from the Two Micron All Sky Survey, which is a joint project of the University of Massachusetts
and the Infrared Processing and Analysis Center/California Institute of Technology, funded by the National
Aeronautics and Space Administration and the National Science Foundation. We are also much obliged for the 
use of the NASA Astrophysics Data System, of the Simbad database (Centre de Donn$\acute{e}$s Stellaires-Strasbourg,
France) and of the WEBDA open cluster database.

\end{document}